\documentclass[10pt,journal,compsoc]{IEEEtran}
\usepackage{graphicx}
\graphicspath{{figures/}}

\usepackage{mathrsfs}
\usepackage{amsmath,amsfonts}
\usepackage{amssymb}
\usepackage{dsfont}
\usepackage{multirow}
\usepackage{subcaption} 
\usepackage{threeparttable}
\usepackage{diagbox}
\usepackage{float}

\ifCLASSOPTIONcompsoc
  \usepackage[numbers]{natbib}
\else
  \usepackage[numbers,compress]{natbib}
\fi
\ifCLASSINFOpdf
\else
\fi
\begin{document}
%
\title{ViKL: A Mammography Interpretation Framework via Multimodal Aggregation of Visual-knowledge-linguistic Features}
%
%
%
%

\author{Xin~Wei,
        Yaling~Tao,
        Changde~Du, 
        Gangming~Zhao,\\
        Yizhou~Yu,~\IEEEmembership{Fellow, IEEE},
        and~Jinpeng~Li,~\IEEEmembership{Member, IEEE}
\IEEEcompsocitemizethanks{
\IEEEcompsocthanksitem Wei, X is with the Ningbo Institute of Life and Health Industry, UCAS, Ningbo 315000, China.
\IEEEcompsocthanksitem Tao, Y, and Li, J are with School of Automation Science and Engineering, South China University of Technology, Guangzhou 510641, Guangdong, China.
\IEEEcompsocthanksitem Du, C is with the Research Center for Brain-Inspired Intelligence,
State Key Laboratory of Multimodal Artificial Intelligence Systems, Institute
of Automation, Chinese Academy of Sciences, Beijing 100190, China.
\IEEEcompsocthanksitem Zhao, G and Yu, Y are with the Department of Computer Science, University of Hong Kong, Pokfulam, Hong Kong.}
\thanks{This work was supported by the National Natural Science Foundation of China under Grant 62106248. Coprresponding authors: Jinpeng Li (jinpeng.li@ieee.org) and Yizhou Yu (yizhouy@acm.org).}
}

%
%

\markboth{Journal of \LaTeX\ Class Files,~Vol.~14, No.~8, August~2015}%
{Shell \MakeLowercase{\textit{et al.}}: Bare Demo of IEEEtran.cls for Computer Society Journals}
\IEEEtitleabstractindextext{%
\begin{abstract}
Mammography is the primary imaging tool for breast cancer diagnosis. Despite significant strides in applying deep learning to interpret mammography images, efforts that focus predominantly on visual features often struggle with generalization across datasets. We hypothesize that integrating additional modalities in the radiology practice, notably the linguistic features of reports and manifestation features embodying radiological insights, offers a more powerful, interpretable and generalizable representation. In this paper, we announce MVKL, the first multimodal mammography dataset encompassing multi-view images, detailed manifestations and reports. Based on this dataset, we focus on the challanging task of unsupervised pretraining and propose ViKL, a innovative framework that synergizes \textbf{Vi}sual, \textbf{K}nowledge, and \textbf{L}inguistic features. This framework relies solely on pairing information without the necessity for pathology labels, which are often challanging to acquire. ViKL employs a triple contrastive learning approach to merge linguistic and knowledge-based insights with visual data, enabling both inter-modality and intra-modality feature enhancement. Our research yields significant findings: 1) Integrating reports and manifestations with unsupervised visual pretraining, ViKL substantially enhances the pathological classification and fosters multimodal interactions. 2) Manifestations can introduce a novel hard negative sample selection mechanism. 3) The multimodal features demonstrate transferability across different datasets. 4) The multimodal pretraining approach curbs miscalibrations and crafts a high-quality representation space. The MVKL dataset and ViKL code are publicly available at \textit{https://github.com/wxwxwwxxx/ViKL} to support a broad spectrum of future research.

\end{abstract}

}

\maketitle

\IEEEdisplaynontitleabstractindextext

%
\IEEEpeerreviewmaketitle

\IEEEraisesectionheading{\section{Introduction}\label{sec:introduction}}

%
%
%
%
\IEEEPARstart{B}{rest} cancer is the foremost cause of morbidity and mortality among women. Timely detection, intervention, and treatment are critical to improving the prognosis, as emphasized in existing literature \cite{waks2019breast}. Mammography stands as the principal tool for early breast cancer screening. Yet, the early indicators of breast cancer are subtle and subject to varying interpretations by different radiologists, presenting challenges in screening accuracy. In the past five years, the advent of deep learning in medical image analysis has sparked a wave of innovation. Models such as convolutional neural networks (CNN) \cite{ting2019convolutional, yap2017automated} and transformers \cite{he2022deconv, mo2023hover} have been increasingly applied for the pathological classification of mammography images. However, these models primarily rely on image data, creating a disconnect from the nuanced clinical practices of radiologists. This gap highlights the need for more integrated and clinically-aligned approaches in breast cancer screening methodologies.

\begin{figure*}
\center
\includegraphics[width=\textwidth]{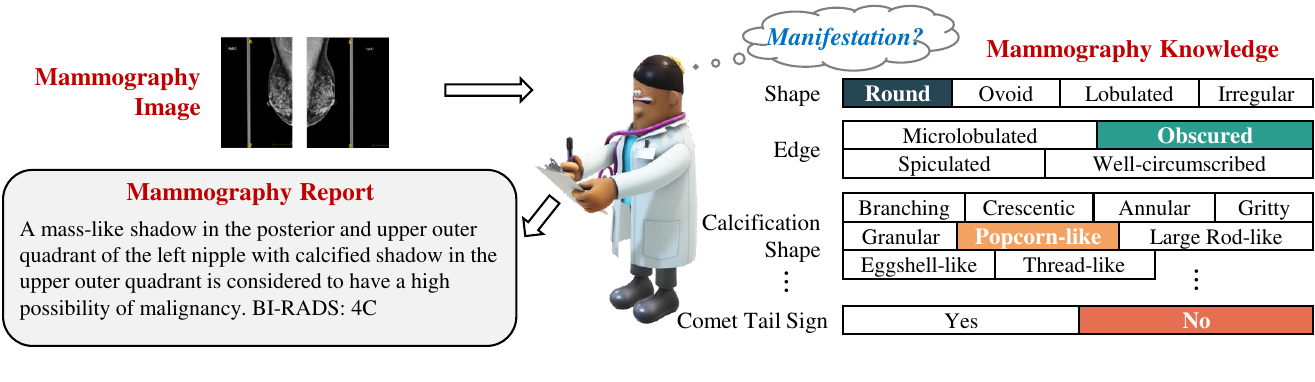}
\caption{Knowledge-guided diagnostic protocol of radiologists for breast cancer. They identify suspected lesions on \emph{mammography images} and refer to the consensus of radiology and their experience to analyze the lesion from the perspective of \emph{manifestations}, e.g., shape, margin and density. Based on these impressions, radiologists write concise and conclusive \emph{mammography reports}. This process inspires the visual pretraining paradigm of ViKL, which aggregates visual, knowledge and linguistic features to build intelligent machines for mammography analysis.} \label{fig:knowledge}
\end{figure*}

Fig.~\ref{fig:knowledge} illustrates the workflow of radiologists in assessing suspected breast lesions. This process begins with the radiologist drawing on both the consensus in radiology and their experience. They observe the lesion from various perspectives, focusing on its manifestations such as shape, contour, and density. The findings are then documented in mammography reports. This workflow encompasses three key elements: 1) \emph{Mammography Images} providing visual information. However, the disease-related features in these images are often complex and subtle, making accurate interpretation challenging. 2) \emph{Manifestations} representing the expert knowledge in analyzing breast lesions, which are rich in dense, semi-semantic information. 3) \emph{Mammography Reports} standing for the linguistic summaries of the radiologist's observations with containing highly sparse and abstract semantic information. The reports encapsulate the radiologist's final interpretation and conclusions based on the visual and manifestation data.

To align computer-aided diagnostic systems to the diagnostic protocol of experts, some studies have investigated the role of reports or manifestations in medical image analysis. The LIDC-IDRI dataset has seven manifestations, e.g., pulmonary consolidation and atelectasis to facilitate the benign/malignant classification of pulmonary nodules \cite{armato2011lung}. However, few manifestations were considered, and manifestations specific to pulmonary nodules are difficult to apply to the mammography interpretation. The ChestX-ray14 dataset has 14 disease labels mined from the X-ray reports via NLP techniques, which shows the possibility of extracting manifestations in reports \cite{wang2017chestx}. In contrast, mammography reports are very succinct, and the diagnostic manifestations are not fully reflected in the report. For example, a manifestation's absence in the report does not imply its absence in the image. Therefore, the keywords extracted from the report are far from representing the expertise of radiologists. To achieve a solid multimodal learning, an exhaustive, generic, and systematic set of manifestations for mammography should be developed and annonated.


The past three years have witnessed a surge in multimodal pretraining algorithms, which have aimed to augment visual features with corresponding linguistic data for various tasks \cite{mogadala2021trends, chen2020uniter, cui2021rosita, radford2021learning, li2021align, yang2022vision, zhou2022generalized}. In medical imaging analysis, most studies have focused solely on radiology-related modalities (such as imaging and reports) for multimodal self-supervised pretraining. For example, REFERS \cite{zhou2022generalized} leveraged supervision signals from X-ray reports and employed self-supervised contrastive learning to develop joint representations. Demonstrating exceptional performance under limited supervision, REFERS even outperforms approaches reliant on structured labels manually added to X-rays. However, REFERS does not incorporate manifestations, which are rich in dense semantic information. We posit that the role of manifestations warrants independent and systematic investigation. In this paper, we aim to systematically formulate, annotate, and explore next-generation multimodal pretraining paradigms tailored to mammography interpretation based on the manifestations.

This work contributes in four aspects.

1) We introduce ViKL, a new self-supervised multimodal pretraining paradigm integrating visual, knowledge, and linguistic features. Using triple contrastive learning, ViKL aggregates inter-modality features among paired data and enhances intra-modality feature alignment. To resolve the issue of semantically similar reports or manifestations resulting in inappropriate negative samples, we have exploited engineering solutions including label smoothing and manifestation deduplication.

2) We propose a novel hard negative sample selection method based on manifestations, termed ManiNeg. It estimates the hardness of negative samples through the Hamming distance between manifestations, effectively addressing the issues of limited sampling space and mismatch between representation and semantics, which exist in traditional representation-based hard negative sample selection methods.

3) We contribute MVKL, the first trimodal mammography dataset, which includes images, manifestations, and reports. This dataset uniquely features a systematically formulated and annotated mammography manifestations.

4) Extensive experiments on MVKL and two public datasets show that the integration of knowledge and linguistic information with visual pretraining significantly improves pathological classification and demonstrates strong cross-dataset generalization. In-depth analysis reveals ViKL's capability in detecting subtle lesions and its generation of high-quality representation spaces.

The rest of this paper is organized as follows. Section 2 reviews relevant works to ViKL. Section 3 introduces the MVKL dataset. Section 4 presents the ViKL framework and optimization method. Section 5 and Section 6 include experimental results with in-depth analysis. Section 7 summarizes take-away information, with limitations and future directions for follow-up studies.

\begin{figure*}
\center
\includegraphics[width=\textwidth]{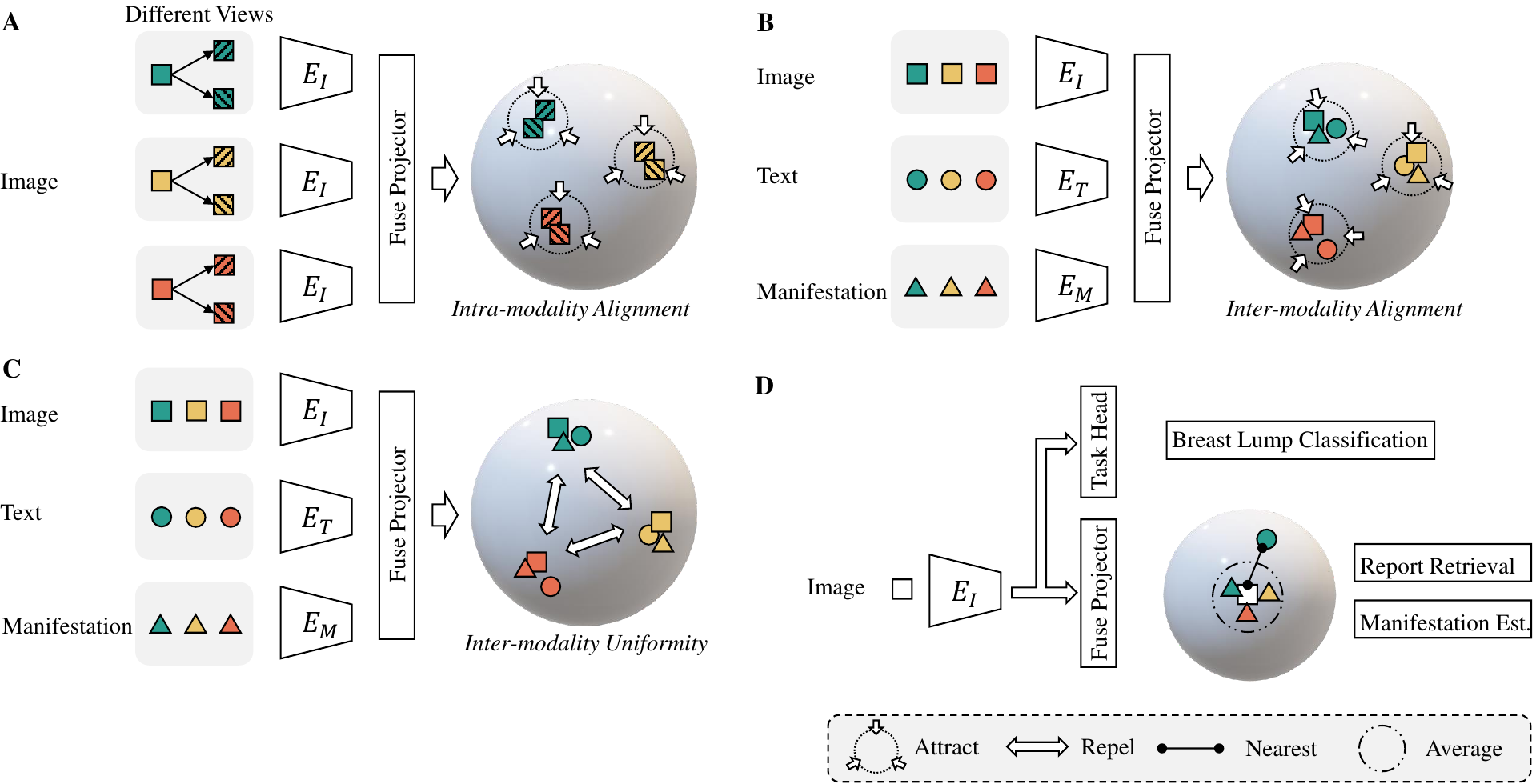} 
\caption{The training (A-C) and test phases of ViKL. \textbf{A}. Mammography views are projected into a 128-dimensional hypersphere embedding space, normalized with $\mathcal{L}_2$ norm. Features from different views are attracted to each other, facilitating intra-modality alignment. \textbf{B}. Data from the three modalities are projected into the same embedding space using respective encoders. Matched modalities of the same instance are attracted to each other, ensuring inter-modality alignment. \textbf{C}. Features from distinct instances are repelled to enhance uniformity across the hypersphere, preserving information effectively. \textbf{D}. The image encoder is capable of improved pathological classification of breast lumps through a simple task head. ViKL model is also versatile, suitable for multimodal tasks like image-report retrieval and manifestation estimation.}\label{fig:flowchart}
\end{figure*}

\section{Related Works}

\textbf{Self-supervised Learning (SSL)}. SSL operates without target task annotations, positioning it within the domain of unsupervised learning. It relies on pretext tasks, which can vary in nature, to provide supervisory signals. The task may involve predicting the rotation angle of a rotated image \cite{komodakis2018unsupervised}, reconstructing the original image from partially masked regions using an auto-encoder (AE) \cite{chen2019self, zhou2019models, zhou2021models}, or restoring the original order of randomly shuffled image patches \cite{noroozi2016unsupervised, zhu2020rubik}. After training, the model’s feature extraction part is retained while the pretext task-specific head is discarded. Subsequently, a simpler head, like a linear classifier, is trained on this feature extractor for the target task. A distinct category within SSL is \emph{instance discrimination}, which treats each instance as a unique class, leading to the emergence of contrastive learning exemplified by SimCLR \cite{chen2020simple} and MOCO \cite{he2020momentum}. In this framework, an image is augmented through rotation, flipping, colorization, etc., and in the embedding space, distances between augmentations of the same image (positive pairs) are reduced, while those between different images (negative pairs) are increased. MOCO uniquely approaches this as a dictionary look-up task, akin to identifying positive matches of a query in a vast dictionary. The effectiveness of SSL hinges on the alignment between the pretext and target tasks, a key determinant of performance.

\noindent{\textbf{Vision-language Pretraining (VLP)}. VLP involves two stages: \emph{pretraining} and \emph{testing}. A prime example of VLP is CLIP \cite{radford2021learning}. During pretraining, CLIP uses image and text encoders to map these modalities into an embedding space, training the encoders to recognize paired images and text. During test, simple classifiers are added to the image encoder to achieve target tasks. CLIP has inspired numerous follow-ups, most utilizing BERT \cite{devlin2018bert} as the language encoder, while adopting various vision encoders. For instance, region-based encoders like Faster R-CNN \cite{ren2015faster} extract regions for VLP, offering high accuracy but increased running time. Conversely, patch-based encoders, such as ViLT \cite{kim2021vilt}, use patch projection, trading a slight decrease in accuracy for reduced running time. Expanding on CLIP, VSE++ \cite{faghri2018vse++} enhances visual-semantic embeddings for cross-modal retrieval by considering the hard negatives. ALBEF \cite{li2021align} aligns image and text representations with a contrastive loss before fusing them via cross-modal attention, facilitating more grounded learning without requiring bounding box annotations or high-resolution images. VLMo \cite{bao2021vlmo} offers a unified vision-language model, leveraging a modular transformer network to jointly learn a dual encoder and a fusion encoder. BLIP \cite{li2022blip} adapts to both understanding and generation tasks in vision-language by bootstrapping web data captions, filtering out noise to enhance learning. CoCa \cite{yu2022coca} combines contrastive and captioning loss in an image-text encoder-decoder foundation model, encompassing both contrastive and generative methods. BEiT \cite{bao2021beit}pretrains vision transformers for masked image modeling tasks, reconstructing original visual tokens from corrupted image patches. PaLI \cite{chen2022pali} integrates visual and textual inputs for text generation, efficiently reusing a large unimodal backbone for both language and vision modeling, thereby transferring existing capabilities and reducing training costs.

\noindent{\textbf{Pretraining in Medical Image Analysis}. The scarcity of annotated data makes pretraining an especially critical approach, generally pursued through three strategies. 1) Some studies opt to pretrain models on general vision datasets like ImageNet. However, the significant disparity between natural and medical images often renders this approach less than ideal. 2) Another method involves seeking free supervisory signals from unannotated data. For instance, Models Genesis utilizes an AE-based SSL approach to reconstruct CT images from their augmented versions \cite{zhou2019models, zhou2021models}, where the encoder is then adapted for downstream tasks. Similarly, our preliminary work with MVCNet leverages multiview contrastive learning to aggregate two-dimensional (2D) views of the same three-dimensional (3D) CT lesion for initial model training \cite{zhai2022mvcnet}. These approaches, however, are limited to single image modalities. ReFs \cite{xie2023refs} integrated self-supervised learning with a supervised reference task, optimizing feature representations for medical image segmentation through gradient matching and dual-task alignment. 3) A third strategy seeks supervision from image reports. For example, REFERS \cite{zhou2022generalized} employs radiography transformers to encode X-ray images and minimizes contrastive alignment loss to pair images with their corresponding free-text reports \cite{zhou2022generalized}. This approach, capitalizing on the complex logic and abstract reasoning in reports, shows promise in supplanting traditional pretraining methods. The existing studies rarely take the manifestation modality into consideration, but we argue that while images are rich repositories of fundamental yet latent disease features and reports encapsulate unstructured, abstract, and semantic information, manifestations stand out as structurable, semi-abstract, and semi-semantic features, serving as a vital bridge between images and reports. Incorporating manifestations into the pretraining process has the potential to yield more relevant and higher-quality representations.

\noindent{\textbf{Manifestations in Medical Image Analysis}. Recognizing the crucial role of domain knowledge, several studies have explored automatic extraction of semi-manifestations. For instance, assessing shape and texture heterogeneity in malignant pulmonary nodules based on segmentation results \cite{xie2018knowledge}, and utilizing manually annotated manifestations for refined modeling. The LIDC-IDRI dataset includes seven attributes for benign/malignant classification of pulmonary nodules. Building on these attributes, various studies have advanced pulmonary nodule diagnosis, such as employing Bayesian frameworks for attribute-based reasoning \cite{zhao2021diagnose} or integrating these attributes into neural network inputs \cite{hussein2017risk}. In our preliminary work, we developed the first tuberculosis dataset featuring seven manifestations and employed a multi-scale feature interaction model for tuberculosis classification and detection \cite{pan2022computer}. Another recent study fused clinical tabular data (like smoking status) and morphometric features (such as ventricle volume) with cardiac MR images to predict myocardial infarction and coronary artery disease risks, leveraging data from the UK Biobank \cite{hager2023best}. Given the distinct radiological manifestations across diseases, it is essential to tailor and meticulously label the manifestation list for each specific disease. 

Additionally, radiomics shares structural characteristics similar to manifestations, making it applicable to multimodal contrastive learning \cite{radio1, radio2}. However, manifestations distinctly differ from radiomics features in their nature and application. While radiomics involves the high-throughput extraction of quantitative image features, 
manifestations offer a more concise and disease-specific approach. They encapsulate higher-level features compared to radiomics, focusing on specific characteristics closely associated with particular diseases. This distinction highlights manifestations as a more targeted and nuanced method in medical imaging, in contrast to the broader, quantitative scope of radiomics.

\noindent{\textbf{Mammography Image Analysis}.} While pathology slides are the gold standard for diagnosing breast cancer, with several deep learning studies focused on their analysis \cite{gecer2018detection, he2020integrating}, mammography remains an essential tool in breast cancer screening. A variety of research has developed multi-scale, multi-view models for detecting breast masses in mammography \cite{cao2019deeplima, liu2021act}. For instance, AGN employed a bipartite graph convolutional network alongside an inception graph convolutional network, enhancing correspondence reasoning by integrating information from ipsilateral and bilateral views \cite{liu2021act}. Also, in light of the limited availability of labeled data, weakly-supervised learning approaches such as multi-instance learning have shown promise in improving sensitivity, utilizing labels of varying granularity \cite{kallenberg2016unsupervised}. 



\section{Dataset}
We have compiled the first trimodal dataset for mammography: The \textbf{M}ammography \textbf{V}isual-\textbf{K}nowledge-\textbf{L}inguistic dataset (MVKL). The dataset consists of 2671 mammography examinations containing breast lumps, along with their imaging reports, manifestations, pathological labels, and segmentation masks for breast lumps. In this section, we will provide a detailed introduction to this dataset.

\textbf{Collection Procedure.} The MVKL dataset comprises data from clinical cases at Ningbo NO.2 Hospital. We first selected mammography examinations with pathology virified benign-malignant evaluations, and utilized their mammography images and corresponding reports as the initialization of the dataset. Based on this, we designed the annotation process for manifestations.

While drafting reports, physicians typically mention only those manifestations they deem significant. Yet, other less emphasized manifestations can also be crucial for representation learning. Therefore, we developed a comprehensive manifestation table to capture a broad spectrum of relevant breast lump traits. The annotation process involved two phases for accuracy. Initially, five attending and chief physicians annotated the breast lumps and their manifestations, referencing the reports and pathology results. In the second phase, a different physician reviewed these annotations, resolving any discrepancies through discussion. This process also included lump segmentation, not utilized in this study but potentially valuable for future research.

Ultimately, this rigorous methodology yielded a collection of 2,764 annotated lumps from 2,671 examinations, complete with corresponding imaging reports and manifestations. Reflecting real clinical case distribution, the dataset aligns closely with actual clinical settings.



\textbf{Mammography Imaging.} In radiological examinations, multiple views are captured to thoroughly represent lesions. For mammography, these typically include four views: left mediolateral oblique (LMLO), left craniocaudal (LCC), right mediolateral oblique (RMLO), and right craniocaudal (RCC). To ensure dataset completeness and versatility for various tasks, we retained all views in their original form, irrespective of the presence breast lumps. However, views without breast lumps are not involved in this study.

\textbf{Manifestations.} We crafted a generic manifestation table based on radiological expertise to describe breast lumps, detailed in Table~\ref{tab:mani}. This table encompasses 8 primary traits of mass and calcification, plus 9 additional traits to characterize breast lumps. Post-annotation, these options are transformed into a binary vector of 35 dimensions to aid neural network interpretation.

\renewcommand{\arraystretch}{1.0}
\begin{table*}[htp]
    \caption{The header of the manifestation table. Options are separated with commas, and slash separators indicate these traits are treated as the same option. The options in the \emph{miscellaneous} row are independent, while the options in other rows are mutually exclusive.}
    \centering
    \begin{tabular}{p{3.5cm}|p{13.7cm}}
    \hline\hline
        \textbf{Manifestations} & \textbf{Options } \\ \hline
        \textbf{mass shape} & irregular, lobulated, ovoid, round  \\ \hline
        \textbf{mass edge} & microlobulated, obscured, spiculated, well-circumscribed  \\ \hline
        \textbf{mass density} & low, median, high  \\ \hline
        \textbf{mass size} & $\leq$2cm, 2-5cm, $>$5cm  \\ \hline
        \textbf{calcification shape} & branching, crescentic/annular/gritty/thread-like, granular/popcorn-like/large rod-like/eggshell-like  \\ \hline
        \textbf{calcification size} & coarse, tiny, uneven  \\ \hline
        \textbf{calcification density} & low, high, uneven  \\ \hline
        \textbf{calcification distribution} & scattered, clustered, linear/segmental  \\ \hline
        \textbf{miscellaneous} & architectural distortion, focal asymmetrical density, duct sign, comet tail sign, halo sign, focal skin thickening/retraction, nipple retraction, abnormal blood vessel shadow, abnormal lymph node shadow \\ \hline\hline

    \end{tabular}
    \label{tab:mani}
\end{table*}

\textbf{Labels.} The MVKL dataset offers pathologically verified benign-malignant evaluations for all lumps, providing dependable diagnostic labels. The label distribution is summarized in Table~\ref{tab:public_dataset}. Besides, accompanying reports include BI-RADS categories assigned by radiologists. The Breast Imaging Reporting and Data System (BI-RADS) categorizes breast images into seven risk levels: 0) inconclusive, 1) negative, 2) benign finding(s), 3) probably benign, 4) suspicious abnormality, 5) highly suggestive of malignancy, and 6) proven malignancy. These categories reflect radiologists’ assessments based on imaging alone, akin to the reports. 


\textbf{Privacy Statement.} Upon collection, patient-identifiable data was promptly removed. This includes specific details and identifiers in the DICOM headers such as patient and physician names, birth dates, institution names and addresses, as well as various identification numbers like Accession Number, Patient ID, Study ID, Study Instance UID, Series Instance UID, and SOP Instance UID.

\section{Method}

In this section, we will first illustrate the architecture of ViKL, and elaborate on the pretraining objectives of it. Finally, we will summary the pretraining objectives and describe the implementation details.

\subsection{Model Architecture}
ViKL is composed of four key components: an image encoder, a text encoder, a manifestation encoder, and a multimodal fuse projector. Specifically, we utilize ResNet50 \cite{resnet} as the image encoder, extracting its penultimate layer output as the image feature. For text encoding, we employ a BERT \cite{devlin2018bert} encoder. Given that manifestations can be represented as vectors, they are encoded using two linear layers. Additionally, an alignment linear layer is appended to each unimodal encoder, standardizing the output dimensions to 128 in our experiments. For both the manifestation and text branches, an extra dropout layer follows the alignment layer, acting as a feature augmenter akin to data augmentation, with a dropout probability of 0.5.

The multimodal fuse projector is pivotal in integrating features from each modality into a unified representation space. Inspired by SimCLR \cite{chen2020simple}, this projector is structured with two linear layers, both input and output features dimensioned at 128. Furthermore, to enhance interactivity in the representation space, the output features undergo $\mathcal{L}_2$ normalization, positioning them on a hypersphere. The schematic diagram of the network is shown in Fig.~\ref{fig:flowchart}, while a more detailed network structure can be found in Appendix~B.


\subsection{Pretraining Objectives}
We pretrain ViKL with dual objectives: \emph{image multi-view contrastive learning} (IMC) and \emph{image-text-manifestation contrastive learning} (ITM). To establish an effective feature space, specific strategies like \emph{label smoothing} and \emph{hard negative sample selection} are incorporated into the loss functions, tailored to the characteristics of each modality. The subsequent sections will delve into these aspects in greater detail.

\textbf{General Objective.} Before proceeding, it's crucial to understand the foundational concept of the contrastive loss function used in our approach. At its core, contrastive learning facilitates unsupervised feature extraction by drawing positive samples closer and pushing negative samples further apart. In unimodal contrastive learning, positive pairs typically consist of different views of the same instance, while negative pairs are formed from distinct instances. Once we achieve alignment in multimodal representations, this principle can effectively be expanded to include pairs across various modalities.

Drawing insights from existing studies \cite{tian2020contrastive, chen2020simple, he2020momentum, radford2021learning}, we can distill the essence of contrastive learning into a fundamental loss function. For two modalities $ m_1 $ and $ m_2 $, and a mini batch of $ N $ pairs, the contrastive loss for a pair of positive sample pair $ \left(\boldsymbol{z}_{m1}^{i}, \boldsymbol{z}_{m2}^{i}\right)$ is defined as

\begin{equation}
    \mathcal{L}_{p}\left(\boldsymbol{z}_{m_{1}}^{i}, \boldsymbol{z}_{m_{2}}^{i}\right) = -\log \frac{\exp \left(\operatorname{sim}\left(\boldsymbol{z}_{m_{1}}^{i}, \boldsymbol{z}_{m_{2}}^{i}\right) / \tau\right)}{\sum\limits_{\substack{k \in N \\ m \in \{m_{1},m_{2}\} \\ k \neq i }} \exp \left(\operatorname{sim}\left(\boldsymbol{z}_{m_{1}}^{i}, \boldsymbol{z}_{m}^{k}\right) / \tau\right)} ,
    \label{loss_onepair}
\end{equation}
where $ \tau $ represents a temperature parameter, crucial in modulating the loss function's dynamic range. In this study, we utilize cosine similarity (denoted as $ sim(\cdot) $) as the metric for measuring feature similarity. Importantly, $ m_{1} $ and $ m_{2} $ are not restricted to being different modalities; when both refer to the same modality, the equation applies to unimodal contrastive learning. In such cases, $ m_{1} $ and $ m_{2} $ correspond to different views of the same instance. It's noteworthy that $ m_{1} $ and $ m_{2} $ are interchangeable. Thus, for a given pair's loss function, the cumulative loss for a mini batch can be obtained through summation:

\begin{equation}
    \mathcal{L}_{cl}\left(\boldsymbol{z}_{m_{1}}, \boldsymbol{z}_{m_{2}}\right) = \sum\limits_{\substack{i \in N \\ n_1 \in \{m_{1},m_{2}\} \\n_2 \in \{m_{1},m_{2}\}\\ n_1 \neq n_2 }}\mathcal{L}_{p}\left(\boldsymbol{z}_{n_{1}}^{i}, \boldsymbol{z}_{n_{2}}^{i}\right).
    \label{loss_base}
\end{equation}

The following sections will delve deeper into the application of these contrastive learning principles across the three modalities in our study, employing these loss functions.
 
\textbf{Image Multi-view Contrastive Learning} is designed to develop a representation space that effectively extracts meaningful image representations. For mammography images, we leverage the contrastive loss to amplify the similarity within ipsilateral breasts and increase separability between patients from different datasets.

We illustrate the process with two views (represented by $ \boldsymbol{v}_{cc} $ and $ \boldsymbol{v}_{mlo} $, e.g., RCC and RMLO in the right breast). Following previous works \cite{chen2020simple, tian2020contrastive}, we extract the feature vector $ \boldsymbol{y} $ for each view using the image encoder $ f_{I} $. i.e., $ \boldsymbol{y}_{cc} = f_{I}(\boldsymbol{v}_{cc}) $ and $ \boldsymbol{y}_{mlo} = f_{I}(\boldsymbol{v}_{mlo}) $. Then, we map the feature vectors to the representation $ \boldsymbol{z} $ by the multimodal fuse projector $ g $, i.e., $ \boldsymbol{z}_{cc} = g(\boldsymbol{y}_{cc}) = g(f_{I}(\boldsymbol{v}_{cc})) $ and $ \boldsymbol{z}_{mlo} = g(\boldsymbol{y}_{mlo}) = g(f_{I}(\boldsymbol{v}_{mlo})) $. By substituting the representation into Eq.~\eqref{loss_base}, we can obtain the basic form of IMC loss function, where $ I $ represents the image modality. 

\begin{equation}
\mathcal{L}_{IMC}^{basic}\left(\boldsymbol{z}_{I}\right) = \mathcal{L}_{cl}\left(\boldsymbol{z}_{cc}, \boldsymbol{z}_{mlo}\right).
\label{loss_imc}
\end{equation}

\textbf{Image-Text-Manifestation Contrastive Learning} extends the scope of image contrastive learning to encompass additional modalities. By aligning the representations of image, text, and manifestations, it enables the seamless application of the unimodal image multi-view loss function to multiple modalities:

\begin{equation}
\begin{split}
    &\mathcal{L}_{ITM}^{basic}\left(\boldsymbol{z}_{I}, \boldsymbol{z}_{T}, \boldsymbol{z}_{M}\right) = \\ 
    &\indent\frac{1}{3}\left( \mathcal{L}_{cl}\left(\boldsymbol{z}_{I}, \boldsymbol{z}_{T}\right) +  \mathcal{L}_{cl}\left(\boldsymbol{z}_{I}, \boldsymbol{z}_{M}\right) + \mathcal{L}_{cl}\left(\boldsymbol{z}_{T}, \boldsymbol{z}_{M}\right)\right),
\end{split}
\label{loss_itm}
\end{equation}
where $I$, $T$, $M$ represent the image, text, and manifestation modalities, respectively. For images, a specific view (e.g. $cc$ or $mlo$) is randomly selected. The processes to derive $\boldsymbol{z}_{T}$ and $\boldsymbol{z}_{M}$ mirror that of the image modality and are therefore not reiterated here. A weighting factor of 1/3 is introduced to balance the contributions of unimodal and multimodal objectives. 
A notable aspect of multimodal contrastive learning is the typical avoidance of selecting negative sample pairs within the same modality \cite{radford2021learning, hager2023best}, implying that $m$ in Eq.~\eqref{loss_onepair} would be limited to $m_{2}$. In this framework, such pairs are invariably negative samples. However, we retain this term, as in a data-limited context, it can aid in achieving a more uniform distribution of representations, thereby preserving information more effectively \cite{wang2020understanding}.

While we have established the basic form of these loss functions, it is crucial to acknowledge that different modalities exhibit varying densities of information, resulting in distinct correspondence levels between them. To address these nuances and ensure model convergence, additional mechanisms are integrated into the objective functions. These will be elaborated upon in the forthcoming sections.

\textbf{Label Smoothing.} In contrastive learning, our goal is to establish a correlation between images and their corresponding reports, but without excessively emphasizing this correlation to the point of pushing away semantically similar reports. This scenario relates to the issue of over-confidence in neural networks, as noted in \cite{guo2017calibration}. Models trained on hard targets tend to assign excessive confidence to their predictions. Similarly, in contrastive learning, which optimizes inter-sample relationships via cross-entropy, there's a risk of over-emphasizing correlations and exhibiting over-confidence. To address this, we integrate label smoothing into the loss function for text, a method proven effective in mitigating over-confidence in the literature \cite{muller2019does, szegedy2016rethinking}. 

Label smoothing works by softening classification targets, thereby reducing over-confidence. It modifies the basic contrastive loss function (Eq.~\eqref{loss_onepair}), which is derived from the cross-entropy loss function. The original ground truth label distribution in this loss function is defined as follows:

\begin{equation}
    P\left(i, k\right) = \left\{\begin{aligned}
        1,i=k\\
        0,i \neq k
    \end{aligned} \right. \qquad i,k \in N,
    \label{p_normal}
\end{equation}

With label smoothing, we introduce a hyperparameter $\alpha \in [0,1]$. This parameter adjusts the ground truth label distribution, effectively softening it to:
\begin{equation}
    P^{ls}\left(i, k\right) = (1-\alpha)P\left(i, k\right) + \alpha/N.
    \label{p_ls}
\end{equation}

By implementing label smoothing, we aim to strike a balance in the correlation between images and texts, enhancing the model's ability to learn nuanced distinctions without generate overly confident predictions.


\textbf{Hard Negative Sample Selection.} The selection of negative samples is critical for the success of contrastive learning. For an anchor sample $\boldsymbol{z}^a$ and its set of negative samples $\boldsymbol{z}^- \in \{\boldsymbol{z}^i_m| i \neq a \forall m\}$, the standard contrastive loss function selects negatives uniformly, which often leads to optimization dominated by more pronounced features. However, in mammography, breast lumps typically occupy a small area and may lack distinct boundaries, rendering their features subtle. To effectively extract this information, contrastive learning must focus on these subtle differences. Thus, it's essential to choose negative samples that are similar yet distinct from the anchor at the lesion level.

However, selecting lesion-level negative samples is not straightforward. Current work on hard negative samples, such as the cross-entropy contained in Eq.~\eqref{loss_base}, or \citet{robinson2020hard}, typically introduces two assumptions. The first is that the representation of samples will strictly align with their semantics, which allow the representation to be utilized as the basis for hard negative sample selection. The second is that there will always be ideal hard negative samples in the mini batch. However, for mammography, due to the small and concealed features of lesions, the representations during training often cannot strictly align with semantics, which can lead to the selection of suboptimal hard negative samples, thereby harming model optimizing. Worse still, the scales of medical datasets often limit batch sizes, which means that ideal hard negative samples may not always be present in the mini batch.

Encouragingly, the situation changed when manifestation is introduced. Manifestation has several crucial advantages. First, manifestation summarizes the core characteristics of breast lumps, meaning it is a reliable proxy for the semantic of mammogram and can be used for hard negative sample selection. Secondly, distance between binary manifestations can be easily measured using Hamming distance, ensuring computational efficiency in the selection of hard negative samples. Lastly, unlike representations that may change with model optimizes, manifestation is an intrinsic attributes of breast lumps and remain constant. This allows for the selection of hard negative samples from the entire dataset rather than just from mini batches.

Based on these observations, we propose a manifestation-centered strategy for sampling hard negative samples, termed as \textbf{ManiNeg}. ManiNeg modifies the default batch sampler, implementing the following batch sampling process: 

1) Randomly select a sample from the training set to serve as the anchor. 

2) Centered on the anchor, with the Hamming distance of the manifestation as the metric, a distribution for negative sampling is constructed. Specifically, we use a truncated Gaussian distribution as the distribution for negative sampling, as detailed in Eq.~\eqref{eq:truncated}.
\begin{equation}
    f(x; \mu, \sigma^2, a,b) = \begin{cases} 
    \frac{\phi(x; \mu,\sigma^2)}{\Phi(b; \mu, \sigma^2) - \Phi(a; \mu, \sigma^2)} , & \text{if } a \leq x \leq b \\
    0, & \text{others}
    \end{cases} 
    \label{eq:truncated}
\end{equation}
In the formula, $x$ represents the Hamming distance between the sample and the anchor's manifestations. $\Phi$ and $\phi$ are the cumulative distribution function and the probability density function of the Gaussian distribution $\mathcal{N}(\mu, \sigma^2)$, respectively. $a$ and $b$ represent the lower and upper bounds of the truncated Gaussian distribution. Specifically, we set the lower bound $a$ to 1 to prevent the anchor itself from being sampled as a negative sample. The upper bound $b$ is determined by analyzing the distribution of Hamming distances within the dataset and is set to 18. $\mu$ and $\sigma$ define the distribution of the hardness of the negative samples, which are treated as hyperparameters.

3) Under this distribution, we will sample $\text{batch size} - 1$ Hamming distances. For each specific Hamming distance, we uniformly sample one from all samples that have that exact Hamming distance to the anchor, and include it in the mini batch.

4) Finally, we will perform deduplication based on manifestations within the batch. Specifically, if multiple samples in the mini batch have the same manifestation, only one sample will be randomly retained. During the training process, every pair of samples in the mini batch will be treated as a negative sample pair. Deduplication helps prevent the model from mistakenly treating potential positive sample pairs as negative ones. 

Additionally, we will incorporate a \textbf{hardness annealing} mechanism during the training process. Hardness annealing gradually increases the hardness of the hard negative samples as training progresses, allowing the model better extract the representations of lesions from easy to challenging. Hardness annealing is implemented by modifying $\mu$ in Eq.~\eqref{eq:truncated} during the training process. For specific settings, please refer to the Appendix~A.

\subsection{Final Objective and Implementations} 
Integrating the mechanisms discussed, the final loss function for ViKL is articulated as follows:

\begin{equation}
\begin{split}
    &\mathcal{L}_{IMC}\left(\boldsymbol{z}_{I}\right) = \mathcal{L}_{cl}\left(\boldsymbol{z}_{cc}, \boldsymbol{z}_{mlo}\right),\\
    &\mathcal{L}_{ITM}\left(\boldsymbol{z}_{I}, \boldsymbol{z}_{T}, \boldsymbol{z}_{M}\right) = \\ &\indent\frac{1}{3}\left( \mathcal{L}_{cl}^{ls}\left(\boldsymbol{z}_{I}, \boldsymbol{z}_{T}\right) + \mathcal{L}_{cl}\left(\boldsymbol{z}_{I}, \boldsymbol{z}_{M}\right) + \mathcal{L}_{cl}^{ls}\left(\boldsymbol{z}_{T}, \boldsymbol{z}_{M}\right)\right),\\
    &\mathcal{L}_{all}\left(\boldsymbol{z}_{I}, \boldsymbol{z}_{T}, \boldsymbol{z}_{M}\right) = \mathcal{L}_{IMC}\left(\boldsymbol{z}_{I}\right) +\mathcal{L}_{ITM}\left(\boldsymbol{z}_{I}, \boldsymbol{z}_{T}, \boldsymbol{z}_{M}\right).
\end{split}
\label{loss_final}
\end{equation}
In these equations, \emph{ls} represents the use of label smoothing. The cumulative function, $\mathcal{L}_{all}$, forms the core of ViKL's pretraining objective. The model underwent pretraining for 300 epochs on a single A100 GPU, totaling approximately 3 hours. For the text branch, we used \emph{bert-base-chinese} \cite{devlin2018bert} as the initial weight. In the image branch, we explored both with and without ImageNet pretrained weights to better understand ViKL's characteristics. Unlike text and manifestations, where no data augmentation is applied at the input stage (though dropout is introduced at the encoder output), image data undergo augmentation. This augmentation, along with the dataset split used in pretraining, aligns with the practices in the downstream tasks phase, which we will discuss in the next section. 



\section{Experiments and Results}

\subsection{Evaluation Protocols}

In our experimental section, we focus on evaluating how ViKL augments the image branch via multimodal contrastive learning, for this is essential clinically. Inspired by CLIP \cite{radford2021learning} and SimCLR \cite{chen2020simple}, we assess the model’s ability to extract features from mammography images using the following protocols:


\textbf{Linear Evaluation (LE)} involves maintaining the pretrained weights of the image branch and adding a trainable linear layer for the classification task. This method evaluates the quality of the extracted features by observing their performance in classification with minimal additional training.

\textbf{Linear Probe (LP)} also keeps the pretrained weights of the image branch fixed. However, classification is performed using a logistic regression classifier, employing the L-BFGS algorithm. This approach allows us to gauge the effectiveness of the extracted features in a standard statistical classification method.

\textbf{Fine-tuning (FT)} uses the pretrained image branch of ViKL as the starting point and further fine-tunes the entire network, including an added linear layer for task-specific purposes. This approach assesses ViKL's adaptability and performance in real-world scenarios.

Both LE and LP limit the number of trainable parameters, thus serving as robust methods to assess the intrinsic quality of ViKL’s features while minimizing the influence of hyperparameters. On the other hand, FT provides insights into the model's practical applicability and performance enhancement in real-world tasks.

\subsection{Experimental Setups}

We now delve into the specific setups of our experiments, providing a clear understanding of how ViKL is evaluated.

\textbf{Downstream Task.} Across all three experimental protocols, we use complete mammography images for classification, identifying the presence of malignant lumps. Given the often indistinct boundaries of breast lumps, this task presents inherent challenges, making it an apt scenario to assess the impact of multimodal pretraining.


\textbf{Label Acquisition.} The labels for these tasks are derived from pathologically validated benign-malignant evaluations, ensuring an objective analysis.

\textbf{Metric.} To ensure consistent and clear comparisons between models, we exclusively use the area under the receiver operating characteristic (AUC). This metric provides a straightforward measure of performance without the complexities introduced by thresholding.


\textbf{Data Augmentation.} The same data augmentation strategy is applied in both pretraining and downstream task phases. This includes \emph{RandomResizedCrop} with a scaling range of (0.5, 1), \emph{RandomHorizontalFlip}, and two layers of \emph{RandAugment} \cite{cubuk2020randaugment}. Notably, as the mammography images are single-channel, color enhancements are not incorporated in \emph{RandAugment}.

\textbf{Dataset Split.} The in-house MVKL dataset is divided into training, validation, and test sets with a 7:1:2 ratio. Care is taken to ensure that different images from the same patient are not included in the same set. Public datasets may follow different splitting schemes, which will be discussed in a subsequent section.

\textbf{Hyperparameters.} Specific hyperparameters for each protocol are listed in Appendix~A.




\subsection{Pulic Datasets for External Evaluation}

In addition to the in-house MVKL dataset, we extend our experiments to two publicly available datasets, CBIS-DDSM \cite{lee2017curated} and INbreast \cite{moreira2012inbreast}, to assess the transferability of ViKL.

\textbf{CBIS-DDSM.} This dataset comprises mammograms specifically annotated for breast cancer diagnosis, including cases marked for calcifications and masses. We adhere to the dataset's provided split scheme, designating 10\% of the training set as the validation set. For CBIS-DDSM, each case comes with a pathology label, which we use directly for classification. 


\textbf{INbreast.} This dataset features a range of mammogram types: normal, those with masses, calcifications, architectural distortions, asymmetries, and multiple findings. We divide this dataset into training, validation, and test sets in a 7:1:2 ratio. INbreast cases are manually categorized into six groups according to BI-RADS standards. For binary labeling, we assign BI-RADS $ \in \{1, 2, 3\} $ as non-malignant and BI-RADS $ \in \{4, 5, 6\} $ as malignant. Notably, BI-RADS 0 is not present in the INbreast dataset.

To ensure comparability, we apply the same experimental protocols to these public datasets as we did with MVKL. The initial weights are pretrained on MVKL to evaluate ViKL's cross-dataset effectiveness. Given the variations in collection years and equipment, these public datasets present mammograms with notable differences in clarity and appearance compared to MVKL, posing a substantial challenge for cross-dataset validation. A detailed summary of these public datasets is provided in Table~\ref{tab:public_dataset}.


\renewcommand{\arraystretch}{1.5}
\begin{table}[htp]
    \centering
    \setlength{\tabcolsep}{6pt}
    \caption{Dataset category distribution. \emph{Ben} and \emph{Mal} are short for \emph{Benign} and \emph{Malignant}.}
    \begin{threeparttable}
    \begin{tabular}{l c c c c c c c c}
        \hline\hline
        Partition & \multicolumn{2}{c}{Training} & &\multicolumn{2}{c}{Validation} & &\multicolumn{2}{c}{Test} \\
        \cline{2-3}\cline{5-6}\cline{8-9}
        Label & Ben. & Mal.& & Ben. & Mal.& & Ben. & Mal. \\
        \hline
        MVKL & 621 & 1325 & &102 & 174 & &158 & 384 \\
        CBIS-DDSM & 1494 & 1083 & &189 & 98 & &428 & 276 \\
        INbreast* & 212 & 75 & & 32 & 9 & & 66 & 16 \\
        \hline\hline
    \end{tabular}
    \begin{tablenotes}
        \footnotesize
        \item[*] BI-RADS 1 category is presented in the INbreast dataset, which typically indicates that no lumps were found in the mammogram. For readability, we merged it into class \emph{Benign}.
    \end{tablenotes}
    \end{threeparttable}
    \label{tab:public_dataset}
\end{table}

\subsection{A Glance on the MVKL Dataset}

Prior to our main experiments, we conducted a quantitative analysis to assess the individual performance of each modality in ViKL when applied to the downstream task. This involved using the untrained encoders of each modality, followed by a linear layer for binary classification. We employed the Adam optimizer \cite{adam} with optimally chosen learning rates to ensure convergence, without additional regularization. The results of these preliminary experiments are detailed in Table~\ref{tab:glance}.


\renewcommand{\arraystretch}{1.5}
\begin{table}[htp]
    \centering
    \setlength{\tabcolsep}{4pt}
    \caption{A Glance on the MVKL Dataset (\%).}
    \begin{tabular}{l c c c c c}
        \hline\hline
         & AUC & Accuracy & Sensitivity & Specificity & F1 \\
        \hline
        Image & $ 68.34 $ & $ 67.59 $ & $ 75.33 $ & $ 48.73 $ & $ 76.72 $ \\
        Text & $ 86.30 $ & $ 70.90 $ & $ 62.60 $ & $ 91.14 $ & $ 75.31 $ \\
        Manifestation & $ 84.84 $ & $ 79.74 $ & $ 89.61 $ & $ 55.69 $ & $ 86.25 $ \\
        \hline\hline
    \end{tabular}
    \label{tab:glance}
\end{table}

Our primary focus is on the AUC results to minimize the impact of classification thresholding. Interestingly, we observed that despite both text and manifestations originating from images, the classification performance using images alone was the weakest. This suggests that neural networks might struggle to extract sufficient useful information from images using simple binary labels. Text and manifestations, in this context, act as refined versions of the original images, potentially guiding the network towards more effective information extraction and representation refinement. This phenomenon echoes findings in natural image domains, as exemplified by CLIP \cite{radford2021learning}, where models pretrained with multimodal contrastive learning showed enhanced robustness compared to those trained solely on ImageNet. 

\subsection{Challenge in Cross-Dataset Generalization}
Supervised learning, due to its simplicity and effectiveness, is a prevalent pretraining approach in deep learning. A typical example is using ImageNet pretrained weights as a starting point for diverse downstream tasks. However, in medical imaging analysis, the limited data volume and label diversity often hinder the generalization capability of models pretrained in a supervised manner. ViKL, with its unsupervised multimodal contrastive learning approach, adeptly overcomes these challenges.

To empirically assess model generalization, we conducted a cross-dataset evaluation using the three datasets. Models were pretrained on each dataset using supervised learning, then tested across all three datasets on a downstream task. We employed the LP protocol for a robust evaluation, with each model beginning pretraining from ImageNet pretrained weights and focusing on benign-malignant classification in both pretraining and downstream phases. The experimental outcomes are detailed in Table~\ref{tab:cross_dataset}, alongside ViKL's results for direct comparison.

The results reveal that while supervised pretrained models show commendable performance on downstream tasks within their original training datasets, their efficacy notably diminishes when applied to different datasets. This underscores the limited generalization capacity of supervised pretraining methods. In contrast, ViKL demonstrates strong, consistent performance across various datasets, indicating its superior ability in generating robust and generalizable representations.

\renewcommand{\arraystretch}{1.5}
\begin{table}[htp]
    \centering
    \setlength{\tabcolsep}{8pt}
    \caption{Cross-dataset evaluation results with the LP protocol. Results are in AUC ($ Mean \pm Std $). Column headers represent the datasets for downstream evaluation, while row headers indicate the datasets for supervised pretraining. The last row includes results for unsupervised ViKL. Italics highlight results with the same pretraining and downstream datasets, while the best performances in cross-dataset scenarios are marked in bold.}
    \begin{tabular}{l l  c c c}
        \hline\hline
        \multicolumn{2}{l}{Dataset} & MVKL & CBIS-DDSM &  INbreast \\
        \hline
        MVKL && \textit{66.51$\pm$0.82} & 61.23$\pm$1.40 & 64.06$\pm$3.22 \\
        CBIS-DDSM && 58.68$\pm$1.96 & \textit{68.98$\pm$0.79} & 64.04$\pm$2.69 \\
        INbreast && 56.74$\pm$1.63 & 59.87$\pm$1.15 & \textit{70.05$\pm$1.79} \\
        \hline
        ViKL  && \textbf{61.24$\pm$1.84} & \textbf{65.58$\pm$0.66} & \textbf{64.11$\pm$0.63} \\
        \hline\hline
    \end{tabular}
    \label{tab:cross_dataset}
\end{table}

\renewcommand{\arraystretch}{1.5}
\begin{table*}[htp]
    \centering
    \setlength{\tabcolsep}{5pt}
    \caption{Comparative AUC percentages (\%) across the MVKL, CBIS-DDSM, and INbreast datasets, expressed as $ Mean \pm Std $. The \emph{IM} model denotes the one pretrained with supervision on ImageNet. \emph{ViKL-R} refers to the variant of ViKL where the image branch is initialized randomly. The \textit{-M} suffix indicates that the model has been fine-tuned on the MVKL dataset, where manifestations are textualized and combined with imaging reports to form the text modality. The abbreviations \emph{LP}, \emph{LE}, and \emph{FT} correspond to the experimental protocols: linear probe, linear evaluation, and fine-tuning, respectively. The top-performing results in each category are highlighted in bold for easy reference.}
    \scalebox{0.91}{
    \begin{tabular}{l c c c   c   c c c   c  c c c  c}
        \hline\hline
         & \multicolumn{3}{c}{MVKL} & &\multicolumn{3}{c}{CBIS-DDSM} & &\multicolumn{3}{c}{INbreast} & \\
         \cline{2-4}\cline{6-8}\cline{10-12}
        Methods & LP & LE & FT && LP & LE & FT &&  LP & LE & FT & Mean \\
        \hline
        IM & 59.64$\pm$1.76 & 71.42$\pm$1.04 & 72.74$\pm$1.51 && 59.97$\pm$1.24 & 67.04$\pm$1.11 & \textbf{75.52$\pm$1.15} && 55.68$\pm$3.32 & 52.08$\pm$8.06 & 73.90$\pm$4.85 & 65.33 \\
        CLIP \cite{radford2021learning} & 51.60$\pm$0.55 & 66.57$\pm$0.66 & 75.70$\pm$1.20 && 57.57$\pm$0.52 & 64.75$\pm$0.17 & 73.02$\pm$1.23 && 52.58$\pm$3.61 & 57.77$\pm$7.52 & 74.62$\pm$5.48 & 63.79 \\
        MedCLIP \cite{wang2022medclip} & 52.90$\pm$0.92 & 65.27$\pm$0.26 & 70.03$\pm$1.12 && 55.90$\pm$1.41 & 61.85$\pm$0.33 & 72.41$\pm$0.75 && 49.10$\pm$2.08 & 59.19$\pm$4.80 & 76.52$\pm$4.57 & 62.57 \\
        ALBEF \cite{li2021align} & 60.93$\pm$2.37 & 68.90$\pm$0.42 & 77.59$\pm$0.33 && 62.08$\pm$1.04 & 67.28$\pm$0.43 & 72.09$\pm$1.00 && 62.03$\pm$2.18 & 62.90$\pm$6.41 & 77.43$\pm$3.03 & 67.91 \\
        BLIP \cite{li2022blip} & 57.80$\pm$0.68 & 70.48$\pm$0.44 & 74.65$\pm$0.27 && 62.55$\pm$0.71 & 66.48$\pm$0.35 & 72.67$\pm$1.56 && 57.65$\pm$6.09 & 71.88$\pm$2.45 & 77.26$\pm$1.91 & 67.94 \\
        Vlmo \cite{bao2021vlmo} & 50.17$\pm$0.21 & 53.56$\pm$2.76 & 68.43$\pm$1.03 && 54.94$\pm$1.37 & 60.01$\pm$0.53 & 68.82$\pm$0.44 && 54.76$\pm$4.76 & 59.33$\pm$4.32 & 61.57$\pm$1.79 & 59.07 \\
        BEiT3 \cite{wang2023image} & 57.46$\pm$1.77 & 70.94$\pm$0.58 & 75.33$\pm$1.40 && 59.05$\pm$0.90 & 62.88$\pm$0.42 & 69.65$\pm$1.53 && 60.18$\pm$1.92 & 69.98$\pm$0.73 & 78.58$\pm$3.86 & 67.11 \\
        MedKLIP \cite{wu2023medklip} & 52.55$\pm$2.09 & 67.14$\pm$0.64 & 70.83$\pm$1.95 && 53.95$\pm$0.77 & 61.35$\pm$0.37 & 70.53$\pm$1.77 && 52.46$\pm$2.85 & 57.33$\pm$3.48 & 69.10$\pm$4.15 & 61.69 \\
        PubMedCLIP \cite{eslami2023pubmedclip} & 60.74$\pm$2.60 & 71.26$\pm$0.53 & 70.41$\pm$4.06 && 61.89$\pm$0.99 & 64.63$\pm$0.32 & 63.72$\pm$1.22 && \textbf{64.96$\pm$5.47}& 66.52$\pm$5.45 & 69.37$\pm$4.40 & 65.94 \\
        BiomedCLIP \cite{biomedclip} & 60.54$\pm$0.73 & 73.62$\pm$0.66 & 75.95$\pm$1.30 && 62.03$\pm$1.00 & 68.54$\pm$0.67 & 68.86$\pm$0.78 && 61.98$\pm$0.20 & 74.34$\pm$3.75 & 74.62$\pm$5.12 & 68.94 \\
        \hline        
        CLIP-M & 52.09$\pm$0.42 & 65.74$\pm$0.95 & 73.48$\pm$2.45 && 57.30$\pm$0.61 & 64.54$\pm$0.58 & 73.03$\pm$1.90 && 52.63$\pm$2.50 & 59.55$\pm$5.27 & 78.79$\pm$3.00 & 64.12 \\          
        MedCLIP-M & 53.20$\pm$1.10 & 65.12$\pm$0.57 & 71.21$\pm$0.78 && 54.70$\pm$0.90 & 62.21$\pm$0.52 & 72.70$\pm$0.88 && 52.32$\pm$3.52 & 58.71$\pm$5.02 & 77.65$\pm$3.31 & 63.09 \\
        ALBEF-M & 61.38$\pm$1.10 & 72.19$\pm$0.39 & 77.88$\pm$0.30 && 60.38$\pm$1.80 & 65.21$\pm$0.26 & 71.43$\pm$1.19 && 61.84$\pm$2.40 & 67.28$\pm$2.45 & 76.49$\pm$2.15 & 68.23  \\    
        BLIP-M & 59.12$\pm$0.44 & 70.55$\pm$0.26 & 75.32$\pm$0.67 && 63.21$\pm$1.01 & 65.76$\pm$0.40 & 72.15$\pm$0.82 && 58.31$\pm$2.52 & 72.15$\pm$3.72 & 76.52$\pm$3.71 & 68.12 \\
        MedKLIP-M & 57.12$\pm$1.04 & 69.43$\pm$0.74 & 74.49$\pm$0.95 && 56.85$\pm$0.97 & 62.55$\pm$0.77 & 73.10$\pm$1.12 && 54.31$\pm$3.12 & 66.53$\pm$4.28 & 72.57$\pm$3.04 & 65.22 \\
        PubMedCLIP-M & 60.36$\pm$1.15 & 72.17$\pm$0.74 & 75.13$\pm$0.68 && 62.16$\pm$0.75 & 65.75$\pm$0.64 & 70.21$\pm$1.01 && 62.15$\pm$4.47& 67.15$\pm$4.43 & 71.53$\pm$3.34 & 67.40 \\
        BiomedCLIP-M & 61.66$\pm$1.10 & 74.15$\pm$0.76 & 76.13$\pm$0.57 && 63.64$\pm$0.68 & 67.98$\pm$0.49 & 71.84$\pm$1.02 && 59.62$\pm$3.46 & 75.15$\pm$3.75 & 78.69$\pm$4.12 & 69.87 \\
        
        \hline
        ViKL-R & 61.16$\pm$2.04 & 72.50$\pm$0.27 & 71.92$\pm$0.49 && 61.46$\pm$1.29 & 66.08$\pm$0.55 & 70.00$\pm$1.30 && 53.22$\pm$1.44 & 61.28$\pm$7.71 & 68.49$\pm$3.93 & 65.12 \\
        ViKL & \textbf{63.81$\pm$1.51} & \textbf{75.14$\pm$0.53} & \textbf{79.06$\pm$0.53} && \textbf{64.59$\pm$1.02} & \textbf{70.94$\pm$0.35} & 72.52$\pm$0.81 && 64.55$\pm$2.49 & \textbf{75.86$\pm$0.81} & \textbf{79.47$\pm$2.15} & \textbf{71.77} \\
        \hline\hline
    \end{tabular}}
    \label{tab:main_results}
\end{table*}

\renewcommand{\arraystretch}{1.5}
\begin{table*}[htp]
    \centering
    \setlength{\tabcolsep}{4pt}
    \caption{Ablation results in AUC percentages (\%) on the MVKL, CBIS-DDSM, and INbreast datasets, presented as $ Mean \pm Std $. The abbreviations \emph{I}, \emph{M}, and \emph{T} correspond to the image, manifestation, and text modalities, respectively. \emph{LS} refers to label smoothing, and \emph{HN} denotes hard negative sample selection. In the \emph{HN} section, $\vartriangle$ indicates the method proposed by \citet{robinson2020hard}, while \checkmark indicates ManiNeg. Other abbreviations used in this table are consistent with those in Table~\ref{tab:main_results}.}
    \scalebox{0.94}{
    \begin{tabular}{c c c c c  c c c  c   c c c   c    c c c c}
        \hline\hline
        &&&&&\multicolumn{3}{c}{MVKL} & &\multicolumn{3}{c}{CBIS-DDSM} & &\multicolumn{3}{c}{INbreast} & \\
        \cline{6-8}\cline{10-12}\cline{14-16}
        I & M & T & LS & HN & LP & LE & FT &&LP & LE & FT &&LP & LE & FT &Mean \\
        \hline
        \checkmark &  &  &  & \checkmark & 55.82$\pm$0.83 & 70.38$\pm$0.96 & 75.45$\pm$0.39 && 62.54$\pm$0.40 & 66.50$\pm$0.13 & 71.00$\pm$0.40 && 50.31$\pm$2.61 & 59.90$\pm$9.03 & 76.30$\pm$1.44 & 65.35 \\
        \checkmark & \checkmark &  &  & \checkmark & 61.69$\pm$1.41 & 71.24$\pm$0.43 & 76.38$\pm$0.26 && 60.80$\pm$0.71 & 65.21$\pm$0.32 & 69.59$\pm$0.98 && 65.08$\pm$2.15 & 63.66$\pm$4.00 & 78.72$\pm$1.44 & 68.04 \\
        \checkmark &  & \checkmark & \checkmark & \checkmark & 59.71$\pm$1.00 & 73.60$\pm$0.33 & 74.80$\pm$0.53 && 62.13$\pm$0.92 & 65.68$\pm$0.32 & 71.80$\pm$1.10 && 50.55$\pm$2.18 & 62.93$\pm$2.80 & 76.11$\pm$1.10 & 66.37 \\
        \hline
        \checkmark & \checkmark & \checkmark & \checkmark &  & 64.93$\pm$1.03 & 75.48$\pm$0.22 & 77.91$\pm$0.53 && 61.83$\pm$0.39 & 67.37$\pm$0.56 & 70.51$\pm$1.10 && 62.73$\pm$0.66 & 69.22$\pm$0.96 & 74.57$\pm$1.60 & 69.39 \\
        \checkmark & \checkmark & \checkmark & & \checkmark & 62.52$\pm$0.45 & 74.89$\pm$0.43 & 77.37$\pm$0.57 && 62.09$\pm$1.03 & 63.08$\pm$0.78 & 69.17$\pm$1.10 && 59.10$\pm$2.21 & 67.93$\pm$3.91 & 80.22$\pm$1.21 & 68.48 \\
        \checkmark & \checkmark & \checkmark & \checkmark & $\vartriangle$ & 61.24$\pm$1.84 & 75.56$\pm$0.44 & 77.46$\pm$0.53 && 65.58$\pm$0.66 & 68.24$\pm$0.49 & 72.48$\pm$1.12& & 64.11$\pm$0.63 & 68.68$\pm$1.78 & 79.16$\pm$3.60 & 70.28 \\
        \hline
        \checkmark & \checkmark & \checkmark & \checkmark & \checkmark & 63.81$\pm$1.51 & 75.14$\pm$0.53 & 79.06$\pm$0.53 && 64.59$\pm$1.02 & 70.94$\pm$0.35 & 72.52$\pm$0.81 && 64.55$\pm$2.49 & 75.86$\pm$0.81 & 79.47$\pm$2.15 & 71.77 \\
        \hline\hline
    \end{tabular}}
    \label{tab:ablation_results}
\end{table*}

\subsection{Downstream Tasks Validation}
In the subsequent sections, we evaluate ViKL's performance on downstream tasks using the previously outlined experimental protocols and setups. To reiterate, we applied LP, LE, and FT protocols across the in-house MVKL, CBIS-DDSM, and INbreast datasets. The findings from these experiments are presented in Tables \ref{tab:main_results} and \ref{tab:ablation_results}.

\subsubsection{Comparison with Supervised Models}

In our analysis, we primarily compare three key results from Table~\ref{tab:main_results}: ViKL, ViKL-R, and IM. ViKL represents our proposed model, where the image branch is preinitialized with ImageNet weights prior to multimodal contrastive learning. In contrast, ViKL-R denotes a variant with a randomly initialized image branch for representation learning. IM serves as a baseline, representing a ResNet50 model pretrained only with supervised learning on ImageNet. This comparison allows us to assess the benefits of multimodal contrastive learning over traditional supervised learning.

The results reveal that ViKL generally outperforms IM across most protocols, highlighting the added value of multimodal learning. Interestingly, ViKL-R, devoid of ImageNet pretraining, attains comparable results to IM, even with a relatively smaller dataset. This indicates that the incorporation of text and manifestation data through contrastive learning effectively enhances the image branch's sensitivity to lesions. This improvement is not only evident in the MVKL dataset used for pretraining but also extends to CBIS-DDSM and INbreast datasets, suggesting the model's adaptability across diverse datasets.


\subsubsection{Ablation Study}

In the ablation study, we explore two crucial aspects: 1) the contribution of each modality, and 2) the effectiveness of the loss function mechanisms (hard negative sample selection and label smoothing) during pretraining in enhancing downstream tasks. The findings are presented in Table~\ref{tab:ablation_results}, demonstrating that these components perform as intended on all the three datasets.

In our modality-related experiments, two findings are particularly notable. 
The first issue relates to the use of only images for contrastive learning, which achieved a baseline performance comparable to the IM model (Table~\ref{tab:main_results}). This indicates that the representation learning did not effectively acquire lesion-related knowledge during pretraining.
We hypothesize that this is because, although contrastive learning can effectively derive image representations, these representations may not necessarily correlate well with lesion characteristics, especially in scenarios with limited data or small lesions.
This phenomenon highlights a crucial consideration in medical image analysis: the potential limitations of applying contrastive learning, successful in natural image domains, might not directly translate to the unique challenges of medical imaging, especially when relying solely on images.


The second key finding in our modality-related experiments involves comparing the combination of images with manifestations against images with text. Introducing text, particularly imaging reports, into contrastive learning is a well-established approach, proven effective in both natural image analysis and medical imaging. However, incorporating manifestations presents a different scenario. While they are fundamental in writing reports and making diagnoses, manifestations are not always explicitly included in reports. Gathering comprehensive manifestations entails some additional annotation effort. Given the emphasis on reducing annotation costs in medical image analysis, justifying such an expense is crucial. Our experiment provides this justification, showing the value of including manifestations in enhancing the learning process.

Finally, we also demonstrate the effectiveness of ManiNeg through ablation study. Table~\ref{tab:ablation_results} shows the results without a specific hard negative sample selection (i.e., weighting hard negatives with cross-entropy) and with the hard negative sample selection method proposed by \citet{robinson2020hard}. Compared to the others, ManiNeg demonstrated the best performance, proving its enhanced capability to extract features from minute lesions.


\subsubsection{Comparison with Off-the-shelf SSL Models}
In the comparative study section, we evaluate ViKL against several notable works: \textbf{CLIP} \cite{radford2021learning}, \textbf{ALBEF} \cite{li2021align}, \textbf{BLIP} \cite{li2022blip}, \textbf{Vlmo} \cite{bao2021vlmo}, \textbf{BEiT3} \cite{wang2023image} for natural image-text pairs, and \textbf{MedCLIP} \cite{wang2022medclip}, \textbf{MedKLIP} \cite{wu2023medklip}, \textbf{PubMedCLIP} \cite{eslami2023pubmedclip}, \textbf{BiomedCLIP} \cite{biomedclip} for medical image-text pairs. Follow the aforementioned protocols, we use the image branches from these models and apply the three experimental protocols (LE, LP, FT) across the three datasets, with results detailed in Table~\ref{tab:main_results}.

The results indicate that comparative methods exhibit less optimal classification performance compared to ViKL. This disparity can partly be attributed to the differing nature of the data used during pretraining. Comparative methods often focus on extracting representations of the salient objects in the images, while ViKL leverages text and manifestations that are finely tuned to describe smaller, localized lumps. This focus enables ViKL to better represent and extract features from small, localized regions, a capability crucial for analyzing challenging lesions in medical imaging.

To demonstrate this, we further pretrained some comparative methods using the MVKL dataset. For a fair comparison, during pretraining, we textualized the manifestation modality (e.g., \textit{mass shape: irregular, mass edge: microlobulated, etc.}), and combined it with imaging reports to form the text modality. The results of downstream task evaluations for such models are marked with a \textit{-M} suffix in Table~\ref{tab:main_results}. It can be observed that after pretraining with MVKL, the models showed improvements in their capability of extracting mammographic representations. ViKL, which utilizes a dedicated manifestation encoder in conjunction with label smoothing and ManiNeg mechanism, demonstrated the best feature extraction capability.

\subsubsection{Comparison with Clinical Radiologists}

To further analyze the significance of ViKL for clinical workflow, we invited experienced clinical radiologists to perform benign and malignant diagnoses on the test set of MVKL and compared their performance with that of ViKL. The specific experimental procedures and results are as follows:

We invited two radiologists specialized in mammograms analysis to participate in this experiment. Similar to the input for ViKL, both radiologists were asked to read single-view mammogram images and determine whether they contained malignant breast lumps. To provide a more diversified comparison, the two radiologists were asked to read the mammograms in different scenarios, henceforth referred to as Radiologist A and Radiologist B. Radiologist A was tasked with judging the nature of the breast lumps as accurately as possible, while Radiologist B was instructed to record results based on first impressions and to make diagnosis about the nature of the breast lumps as quickly as possible. We randomly split the test set into two equal parts and assigned them to the two doctors for diagnosis, with their results recorded in the Table~\ref{tab:clinical}. We also included the results of ViKL at different thresholds for a comprehensive comparison.

\renewcommand{\arraystretch}{1.5}
\begin{table}[htp]
    \centering
    \setlength{\tabcolsep}{3pt}
    \caption{Results of clinical comparison experiments (\%).}
    \begin{tabular}{l c c c c c}
        \hline\hline
         ~ & Time & Threshold  & Sensitivity & Specificity & Accuracy \\ 
         \hline
        Radiologist A & 10.091s & - & 58.20 & 82.05 & 65.17 \\ 
        Radiologist B & 5.935s & - & 39.68 & 97.44 & 56.55 \\ 
        \hline
        ViKL & 0.005s & 0.87 & 40.24 & 96.62 & 56.13 \\ 
        ViKL & 0.005s & 0.74 & 58.50 & 86.12 & 67.76 \\ 
        ViKL & 0.005s & 0.50 & 89.70 & 45.33 & 76.46 \\ 
        \hline \hline
    \end{tabular}
    \label{tab:clinical}
\end{table}

\begin{figure}[htp]
    \centering
    \includegraphics[width=0.45\textwidth]{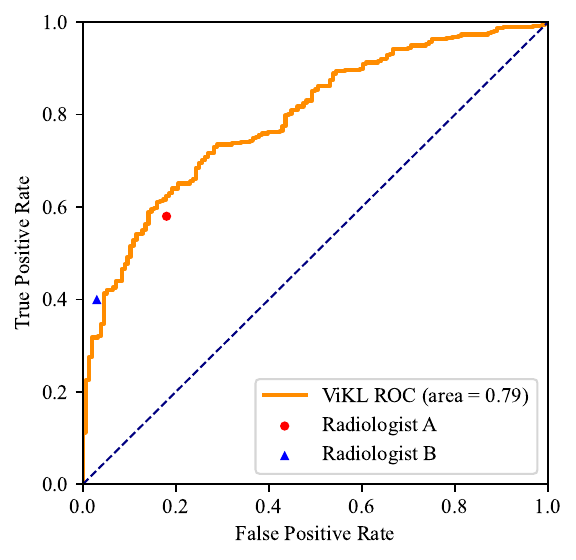}
    \caption{The ROC curve for ViKL, along with the metrics from two radiologists. When diagnosing with single-view mammograms, their classification performance is comparable, but the probabilistic outputs from ViKL provide it with greater flexibility.}
    \label{fig:clinical_roc}
\end{figure}
For a intuitive comparison, we plotted the ROC curve for ViKL and marked the metrics of the two radiologists on it in the Fig.~\ref{fig:clinical_roc}. It is evident that ViKL exhibited classification performance on par with that of the radiologists. Based on this observation, We can analyze the potential implications of ViKL for clinical workflows.

Firstly, unsurprisingly, ViKL possesses a significantly faster reading speed than humans, which offers a distinct advantage in terms of time cost for analyzing large datasets, such as retrospective analyses of historical data. Secondly, diagnostic habits of radiologists are shaped over years of learning and practice, implying a unchanging preference for sensitivity and specificity. For example, in our experiments, radiologists tend to diagnose images as benign in the absence of sufficient evidence, leading to higher specificity and lower sensitivity. Since ViKL outputs probabilities of benignity or malignancy, it offers greater flexibility. Users can adjust the threshold according to their needs, balancing between sensitivity and specificity, thus complementing human diagnostic capability with the advantages of ViKL. Finally, the focus of ViKL is on multimodal representation learning. To intuitively compare the performance of representation learning and reduce the impact of complex network structures, we adopted a straightforward downstream task for evaluating the model: using the complete single-view mammogram to determine whether it contains malignant lumps. This also means that there is significant potential for further improvement in the model's performance. For instance, ViKL could leverage transfer learning techniques to transfer weights to models that involve multiple mammographic views or lump localization, thereby further optimizing the model's usability and performance.

It is noteworthy that diagnosing breast lumps with a single-view mammogram is a challenging task. This challenge arises from factors such as overlapping breast tissues, breast density, and lump size. Meanwhile, the lack of the paired mammogram view further exacerbates these challenges. This is also why currently the diagnostic metrics for both radiologists and ViKL are relatively low. Building on ViKL's demonstrated capabilities in representation learning, as outlined above, enhancing the model's diagnostic performance will be a key focus of our future work.

\subsection{Multimodal Evaluation}

Interactions among modalities represent a cornerstone in multimodal research. The alignment of representations from disparate modalities within a shared space unlocks the potential for complex tasks like retrieval and generation across these modalities. However, the relatively small scale of the MVKL dataset imposes a limitation, resulting in a representation space that is sparse. This sparsity potentially hampers the robustness and practicality of intermodal interactions. Nonetheless, even within this limited representation space, we can still illustrate the existence of meaningful correlations between modalities. This assertion will be further validated through two retrieval-based experiments that we present in the following sections.


\textbf{Manifestation Estimation.} This experiment uses images as inputs to estimate the manifestations of lumps within them, achieved by retrieving neighboring manifestations in the representation space based on the image representations. We first compute the representations of all manifestations in the training set using ViKL. For a given test image representation, we select associated training set manifestations based on their cosine distances from the image representation, mirroring the distance metric used in pretraining.
 
The structured and discrete attributes of manifestations contribute to their more sparse distribution in the representation space, compared to other modalities. To counteract this and achieve robust retrieval outcomes, we implement a distance threshold strategy. This involves selecting nearby manifestations relative to the input image and determining the classification result based on the proportion of positive votes in each dimension. 
Adjusting the classification threshold below this value allows for fine-tuning the balance between sensitivity and specificity. Comprehensive classification metrics, calculated across the 35 manifestation dimensions, are detailed in Table~\ref{tab:mani_est}.

\renewcommand{\arraystretch}{1.5}
\begin{table}[htp]
    \centering
    \setlength{\tabcolsep}{4pt}
    \caption{Average metrics over 35 dimensions of manifestation (\%). For each input image, manifestations whose representations' distances are less than $T_{distance}$ will be averaged, and the dimensions exceed $T_{classification}$ in this averaged vector are regarded as positive.}
    \begin{tabular}{c c c c c}
        \hline\hline
         $T_{distance}$ & $T_{classification}$ & Accuracy & Sensitivity & Specificity \\
        \hline
          0.4 & 0.15 & 72.72 & 73.30 & 72.62  \\
         0.4 & 0.12 & 66.33 & 80.55 & 63.81  \\
         0.4 & 0.25 & 82.80 & 50.21 & 88.57  \\
        \hline\hline
    \end{tabular}
    \label{tab:mani_est}
\end{table}

\textbf{Image-Report Retrieval.} In this experiment, we compute the results by identifying the single nearest neighbor report in the representation space, leveraging the relatively dense distribution of reports. This process involves pre-computing representations for all reports in the training set and then locating the nearest neighbors within this set using image representations from the test set. However, evaluating the accuracy of the retrieved results is challenging due to the specialized medical knowledge required. To address this, we employ a straightforward and objective method: analyzing the consistency of BI-RADS categories in the retrieved reports with those in the ground-truth reports of the corresponding images. BI-RADS evaluations by radiologists are a universal component of mammography reports and thus offer a reliable basis for assessing the effectiveness of the retrieval task.


\begin{figure}[htp]
    \centering
    \includegraphics[width=0.4\textwidth]{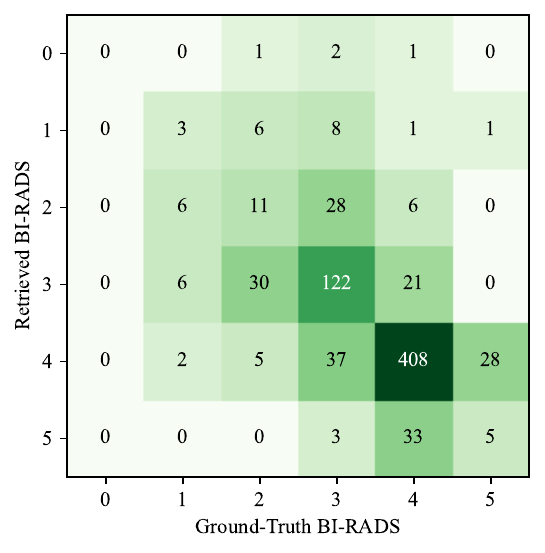} 
    \caption{Confusion matrix of image-report retrieval experiment.}
    \label{fig:cm}
\end{figure}

The confusion matrix is depicted in Fig.~\ref{fig:cm}. Notably, the retrieval results predominantly cluster near the diagonal, indicating that ViKL effectively matches images with reports through multimodal contrastive learning. 

\subsection{ViKL Reduces the Data Requirement in Fine-tuning}
A key goal of pretraining is to alleviate the data constraints often encountered in downstream tasks, where data scarcity and high annotation costs are prevalent. To evaluate this, we deliberately reduced the training set of the MVKL dataset to 20\%, 40\%, 60\%, and 80\% of its full size. We then conducted fine-tuning experiments on these subsets to assess how ViKL and the supervised IM model adapt to varying data availability in downstream tasks. The outcomes of these experiments are illustrated in Fig.~\ref{fig:limited_data}. As anticipated, ViKL demonstrates a markedly lower sensitivity to reductions in training data volume compared to the supervised IM model. This finding underscores ViKL's efficiency in leveraging limited data for training.

Moreover, even during pretraining, ViKL exhibits an ability to reduce data requirements. As discussed earlier, ViKL-R, pretrained on just a few thousand images from the MVKL dataset, achieves a performance level comparable to that of the IM model, which is pretrained on the extensive ImageNet dataset. This indicates that multimodal contrastive learning, by providing nuanced guidance for model optimization, enables more efficient use of limited training data. 

\begin{figure}[htp]
    \centering
    \includegraphics[width=0.45\textwidth]{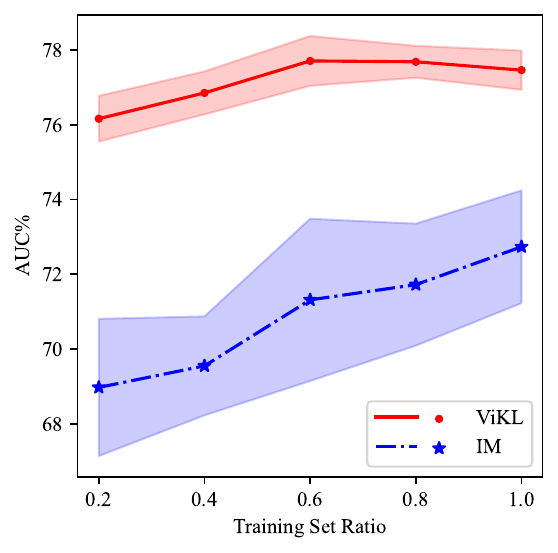}
    \caption{ViKL shows superior adaptability compared to the supervised IM model when fine-tuning on the MKVL dataset with reduced training data, illustrating less performance reduction as the availability of downstream data decreases.}
    \label{fig:limited_data}
\end{figure}


\subsection{ViKL Provides Evidence for Clinical Decision}

In our earlier sections, we discussed how ViKL enhances the image branch's capability for detailed feature extraction by using text and manifestations that describe localized lumps. To directly observe this enhanced feature extraction ability, we utilize the class activation map (CAM) \cite{zhou2016learning}, a common tool for visualizing neural network activations. Specifically, we employ Grad-CAM \cite{selvaraju2017grad} to generate activation maps from the third layer of the ResNet50 model, which has been fine-tuned using ViKL on the MKVL dataset. Additionally, for accurate visualization and comparison, we include markings of lumps as annotated by professional radiologists. The results of this visualization are showcased in Fig.~\ref{fig:vis}. These visualizations demonstrate ViKL's effective localization of small lesions, highlighting its refined ability to capture minute features in medical imaging.

\begin{figure*}[ht]
\centering
\begin{subfigure}{0.20\textwidth}
  \centering
  \includegraphics[width=\linewidth]{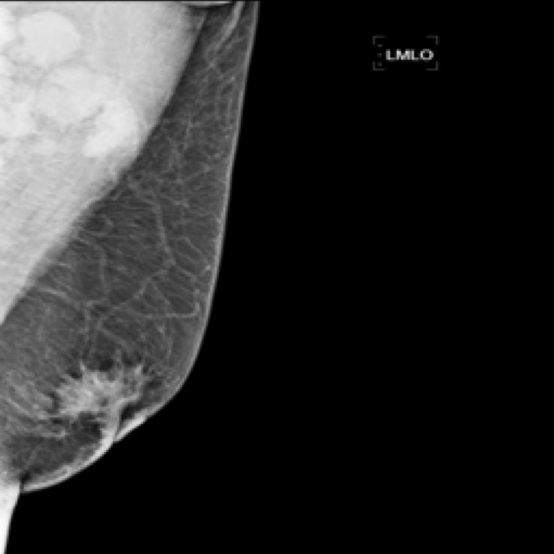}
\end{subfigure}%
\begin{subfigure}{0.20\textwidth}
  \centering
  \includegraphics[width=\linewidth]{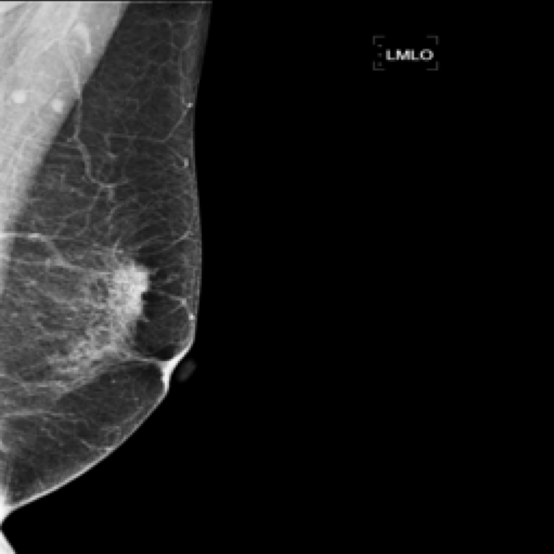}
\end{subfigure}%
\begin{subfigure}{0.20\textwidth}
  \centering
  \includegraphics[width=\linewidth]{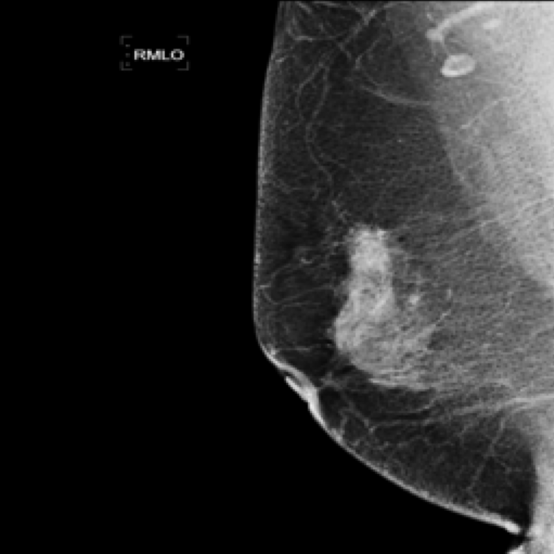}
\end{subfigure}%
\begin{subfigure}{0.20\textwidth}
  \centering
  \includegraphics[width=\linewidth]{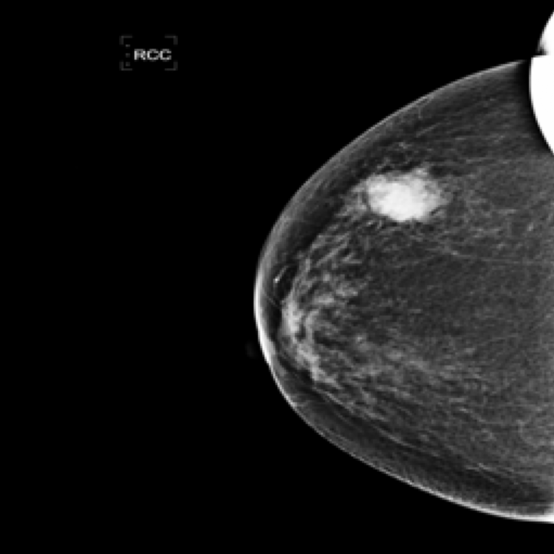}
\end{subfigure}%
\begin{subfigure}{0.20\textwidth}
  \centering
  \includegraphics[width=\linewidth]{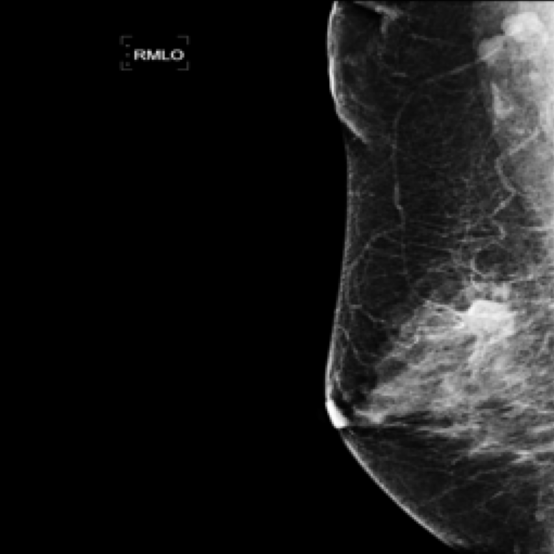}
\end{subfigure}%
\hspace{0mm}
\begin{subfigure}{0.20\textwidth}
  \centering
  \includegraphics[width=\linewidth]{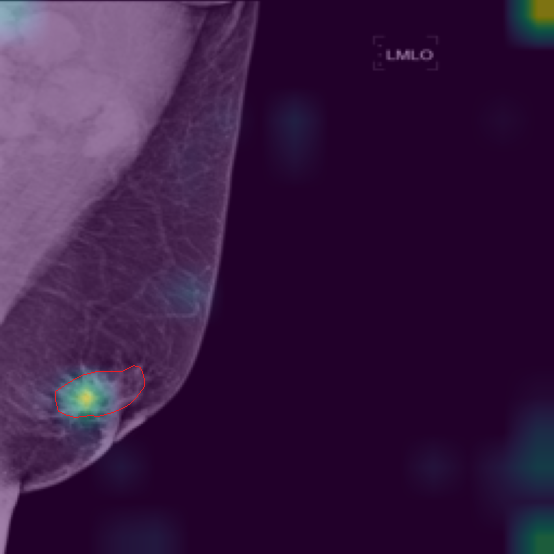}
\end{subfigure}%
\begin{subfigure}{0.20\textwidth}
  \centering
  \includegraphics[width=\linewidth]{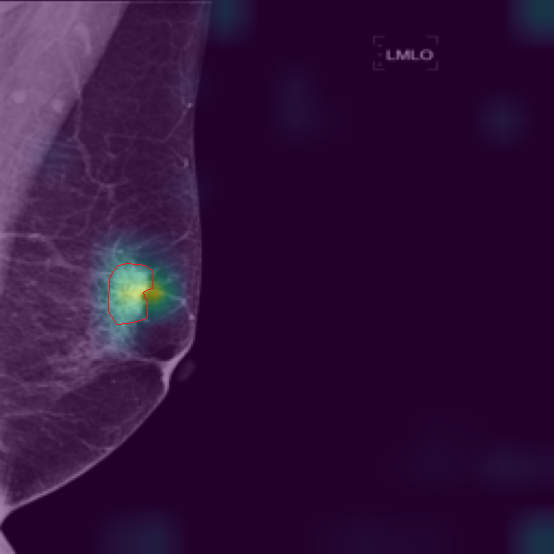}
\end{subfigure}%
\begin{subfigure}{0.20\textwidth}
  \centering
  \includegraphics[width=\linewidth]{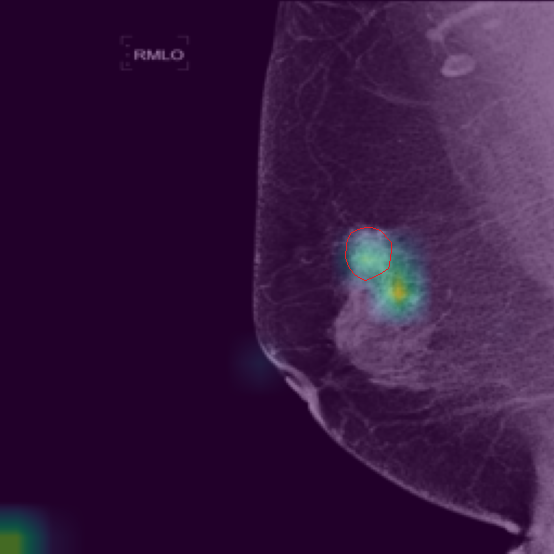}
\end{subfigure}%
\begin{subfigure}{0.20\textwidth}
  \centering
  \includegraphics[width=\linewidth]{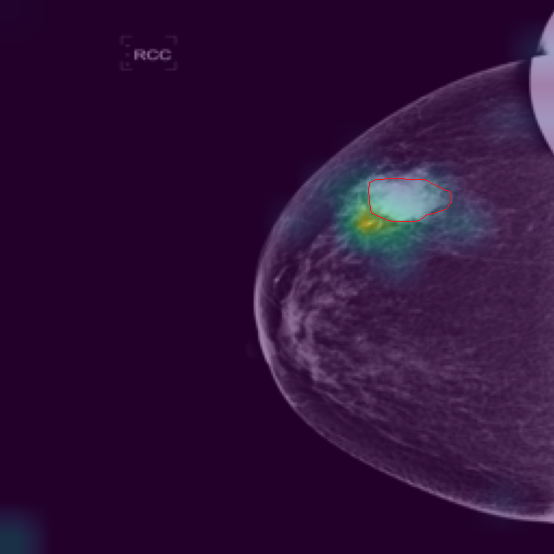}
\end{subfigure}%
\begin{subfigure}{0.20\textwidth}
  \centering
  \includegraphics[width=\linewidth]{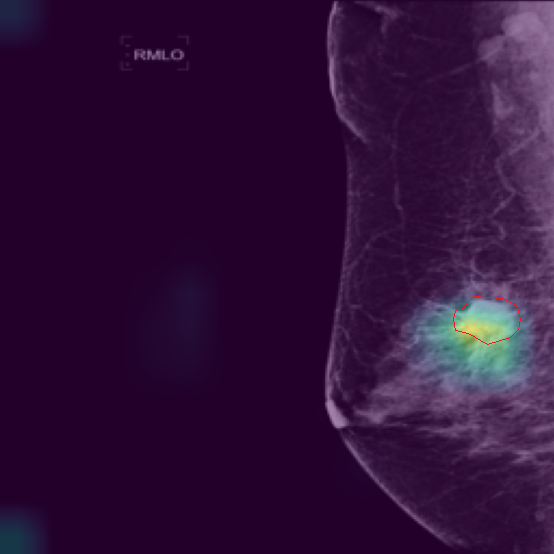}
\end{subfigure}%
\caption{Exemplary Grad-CAM visualizations from ViKL. The first row displays the original mammogram images, while the second row presents the corresponding Grad-CAMs generated from layer 3 in ViKL. In these second-row images, breast lumps, as annotated by professional radiologists, are outlined for clear comparison. These visualizations demonstrate ViKL's capability, achieved through unsupervised multimodal contrastive learning, to precisely localize the breast lump regions within the entire mammogram.}
\label{fig:vis}
\end{figure*}

\subsection{ViKL Enhances Confidence Calibration}

\citet{guo2017calibration} delves into the critical issue of confidence calibration in neural networks, emphasizing that an ideal classifier's predicted confidence should accurately reflect the actual likelihood of a correct prediction. They noted that modern neural networks, despite improved classification metrics, often suffer from miscalibrated confidence estimates. This issue is particularly pertinent in healthcare applications, where users interpret a classifier’s confidence as an indicator of lesion severity and ambiguity.

To address neural network miscalibration, \citet{muller2019does} confirmed the effectiveness and ease of integration of label smoothing. We incorporated label smoothing during ViKL's pretraining phase, primarily to address the challenge of multiple images corresponding to the same report. In this section, we shift our focus to explore how label smoothing contributes to improving model calibration.

Following the method outlined by \citet{guo2017calibration}, we assess calibration by categorizing results into bins based on predicted confidence, and then calculating classification accuracy within each bin. This approach allows us to create \emph{reliability diagrams}, where a line graph plots confidence against accuracy. Additionally, we calculate the expected calibration error (ECE) by weighting the absolute difference between confidence and accuracy in each bin by the bin's sample size. Ideally, in these diagrams, the plotted curve should closely follow the diagonal line.

We compared the calibration performance of three models fine-tuned on the MKVL dataset: ViKL with label smoothing (ViKL w/ LS), ViKL without label smoothing (ViKL w/o LS), and the ImageNet-supervised model (IM). Notably, label smoothing was only applied during pretraining, with no additional implementation during fine-tuning. Our analysis employed 10 bins across the 0.0-1.0 confidence range, excluding bins with fewer than 10 samples for numerical stability. The resulting reliability diagrams and ECEs are illustrated in Fig.~\ref{fig:over-confidence}. The analysis reveals that multimodal contrastive learning inherently improves model calibration. This may stem from the limitations of IM in building a semantically meaningful representation space using one-hot labels in supervised training, which undermines the reliability of confidence scores. Conversely, integrating text and manifestations appears to alleviate this limitation. Further enhancing this, label smoothing refines the representation space, granting ViKL w/ LS superior calibration capabilities.

\begin{figure}[htp]
    \centering
    \includegraphics[width=0.45\textwidth]{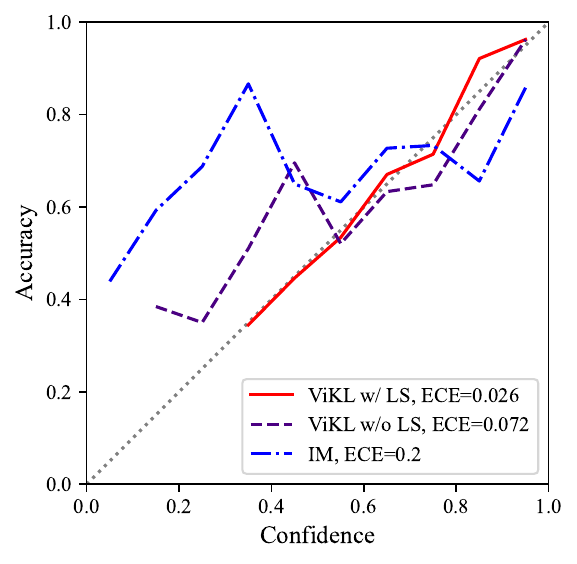} 
    \caption{Reliability diagrams highlighting confidence calibration. Benefiting from the integration of multimodal information and the implementation of label smoothing mechanisms, ViKL demonstrates excellent properties of confidence calibration.}
    \label{fig:over-confidence}
\end{figure}

\begin{figure*}[htp]
  \centering
  \begin{subfigure}[b]{\textwidth}
    \centering
    (a)
    \includegraphics[width=0.95\textwidth]{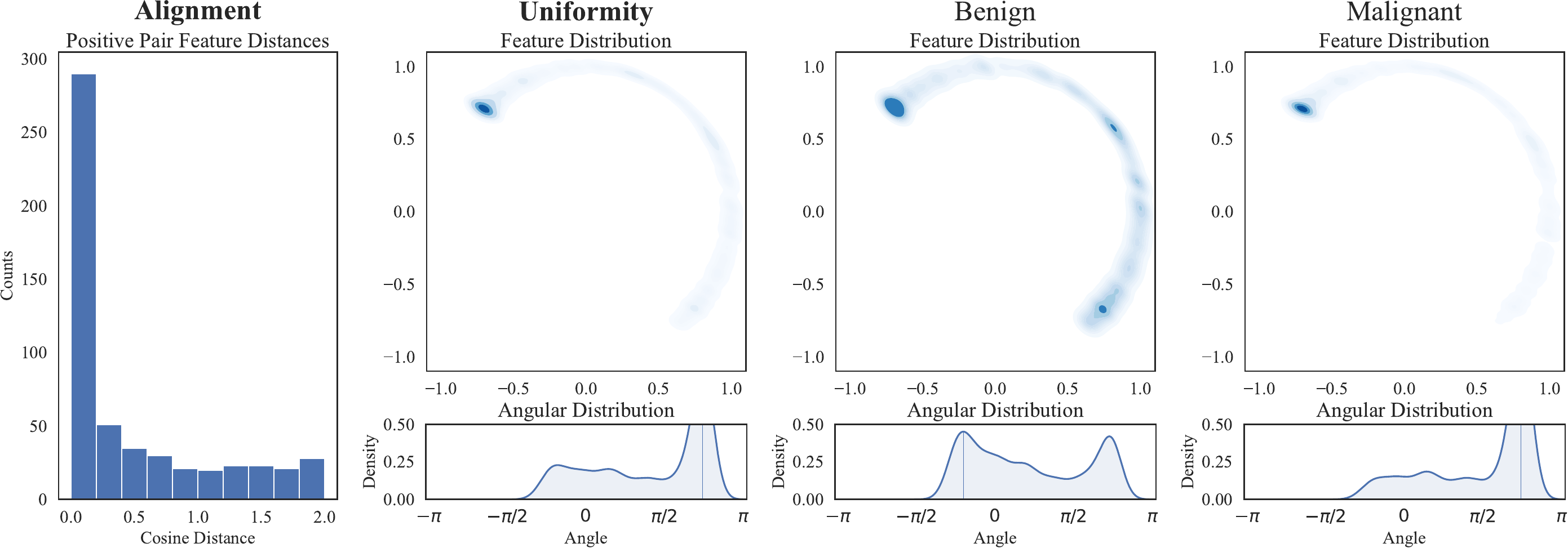}
    \label{fig:ua_im}
  \end{subfigure}
  \begin{subfigure}[b]{\textwidth}
    \centering
    (b)
    \includegraphics[width=0.95\textwidth]{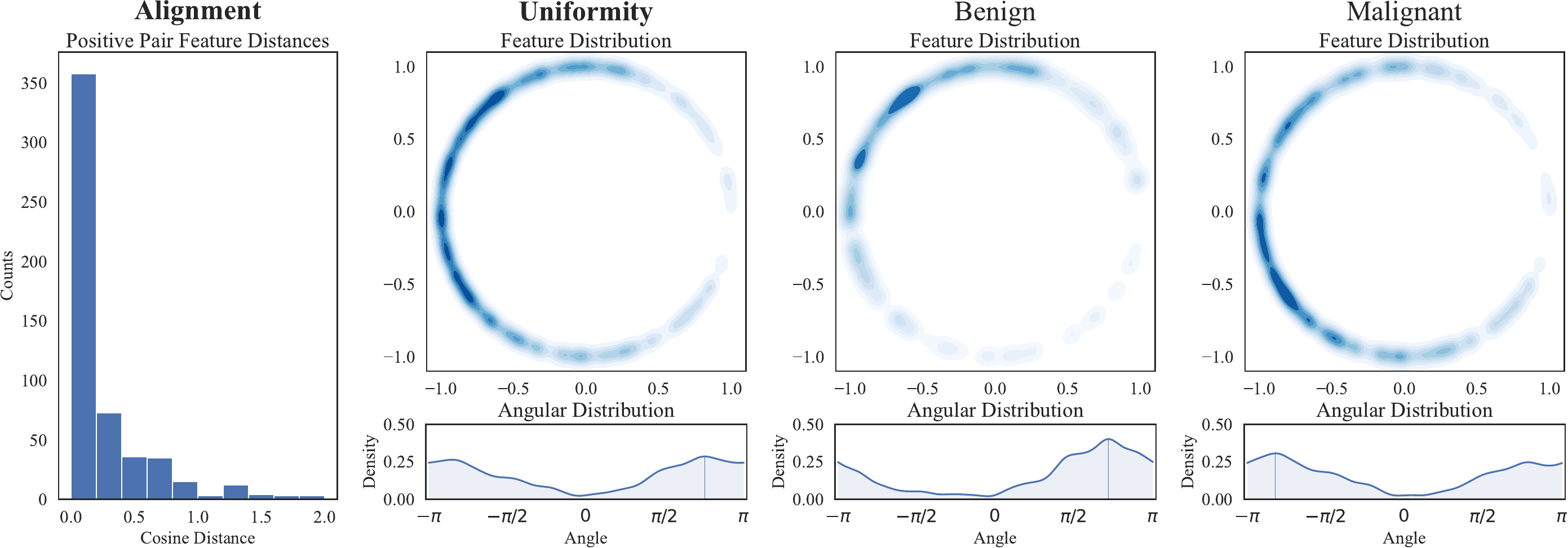}
    \label{fig:ua_vikl}
  \end{subfigure}
  \caption{Alignment and uniformity analyses of the representation spaces. Panel (a) illustrates the model trained via supervised learning, and panel (b) depicts the model trained with multimodal contrastive learning. \emph{Alignment} is demonstrated through histograms representing the distances within positive sample pairs. \emph{Uniformity} is conveyed via kernel density estimation (KDE) of the representation distribution. The planar distribution is visualized using Gaussian KDE, while the angular distribution employs von Mises KDE, with the peak points indicated by vertical lines. Additionally, on the right side of each panel, uniformity analyses for benign and malignant samples are separately presented, allowing for a nuanced comparison of the model's performance across different sample classifications.}
  \label{fig:ua}
\end{figure*}

\subsection{ViKL Crafts A High-quality Representation Space}

In the preceding sections, we explored how ViKL enhances its representation space, particularly from the perspective of calibration. This section delves deeper, focusing squarely on the representation space itself to understand this enhancement. \citet{wang2020understanding} provided a framework to evaluate the quality of a representation space based on two key properties: \textbf{Alignment} and \textbf{Uniformity}. Alignment implies that semantically similar samples should be proximal in the representation space, while uniformity suggests that representations should be evenly dispersed throughout the space to maximize information preservation. Balancing these two, often conflicting criteria, is crucial in the realm of representation learning. 

Adopting the methodology of \citet{wang2020understanding}, we aim for an intuitive understanding of these properties. We configured ViKL's representation dimension to 2 (i.e. $\boldsymbol{z} \in \mathbb{S}^2$) and trained it using both multimodal contrastive and supervised learning methods. For the latter, we used labels from the downstream task. After training with the training set, we used test set images to visualize the representation space. To assess alignment, we generated histograms showing distances among positive sample pairs. For uniformity, we employed kernel density estimation (KDE) to depict the spread of representations: Gaussian KDE for planar distribution and von Mises KDE for angular distribution. These visualizations are presented in Fig.~\ref{fig:ua}.

Multimodal contrastive learning, being the direct training objective, showcases superior alignment. Additionally, its representations are more uniformly distributed across the space, indicating a richer preservation of information from the input images and thus enhanced adaptability and robustness in downstream tasks. Beyond a general visualization of the entire test set, we also conducted separate visualizations for samples classified as benign and malignant. Intriguingly, despite benign-malignant classification not being a direct objective of multimodal contrastive learning, the peak points of their angular distributions are distinct. This observation suggests that the network has indeed learned features pertinent to downstream tasks via text and manifestations, even if these features are subtle.


\section{Discussions}
Experiment results clearly show that the combination of images and manifestations outperforms that of images and text. This suggests that the key improvements in ViKL are largely due to the inclusion of manifestations. Text-based feature extraction, requiring complex models like BERT and a diverse data set, is challenging in the medical field. This often results in text features incorporating noise that can hinder representation learning. Additionally, the issue of a single report corresponding to multiple images, albeit mitigated by label smoothing, still presents a challenge for contrastive learning.

Manifestations effectively address these concerns. Each dimension of a manifestation is closely linked to specific lesion characteristics, reducing potential noise. The issue of multiple images matching the same manifestation can be resolved through deduplication. Remarkably, encoding for manifestations doesn't necessitate complex models like BERT; a simple two-layer linear model is adequate. This aligns with recent research findings, such as those by \citet{hager2023best}, which confirm the effectiveness of combining tabular data with image data for enhanced performance in downstream tasks. Moreover, manifestations are well-integrated into medical practice, as radiologists routinely interpret these features for diagnosis. This suggests that annotating manifestations could be seamlessly incorporated into routine assessments, possibly through a simple checkbox tool. However, depending solely on manifestations has its drawbacks, particularly in limiting the flexibility of information extraction.

In our comparative analysis across three datasets, ViKL consistently outperformed models pretrained on ImageNet. This is particularly noteworthy considering the scale of data used: ViKL was trained on about two thousand samples, while ImageNet pretraining involved million-level labeled images; ViKL operates under an unsupervised learning paradigm, relying solely on pairing information without the need for pathology labels, whereas ImageNet uses labeled images for supervised training. These results not only demonstrate the efficacy of ViKL's training approach but also highlight the importance of domain-specific pretraining for medical imaging tasks. The substantial domain gap between general imagery and medical images suggests that pretraining on similar or related modalities is more beneficial.

While ViKL demonstrates outstanding performance, it is important to recognize its limitations. 1) ViKL necessitates disease-specific manifestation sets and relies on the annotation efforts of radiologists.
2) In radiology, manifestations are defined for specific organs or diseases, which restricts ViKL from being entirely task-agnostic. In an era that foundation models are increasingly prevalent, possessing task-agnostic capabilities is highly advantageous. It facilitates seamless adaptation across diverse datasets and paves the way for the development of general models capable of zero-shot learning across tasks. However, this challenge is not unmanageable. Considering the high information density of tabular data and the finite variety of lesions and organs, the prospect of creating a comprehensive, universal manifestation table for wider applications remains a viable possibility.

Building on the current findings, we foresee several exciting future research directions for ViKL:

1) Manifestations present a fertile ground for future exploration. The prospect of creating a universal, task-agnostic manifestation table and reducing data requirements for manifestations compared to other modalities opens up numerous research opportunities.

2) The effort involved in developing and annotating manifestations is noteworthy. ViKL was trained with just over two thousand complete trimodal data pairs, yet it outperforms models pretrained on ImageNet. There's room for improvement as data volumes increase. Clinically, many image-report pairs exist, and we can infer manifestations from these to expand our dataset. This expansion could further train and enhance ViKL, leveraging clinical data abundance for greater performance.

3) While our primary focus has been on unimodal image performance for diagnosis, the potential applications of multimodal pretrained models are vast and extend well beyond this scope. We can delve into more intricate multimodal tasks, such as inter-modality generation and retrieval. Particularly intriguing is the possibility of generating more comprehensive, reliable, and information-dense reports by transitioning from images to manifestations, and ultimately to detailed reports.

4) Although ViKL has showcased implicit capabilities in lump localization with full mammography images as input, there is significant scope to extend these capabilities to explicit localization through specialized task heads. The current focus has been on benign-malignant screening tasks in breast lumps. However, with the MVKL dataset providing detailed lump location labels and segmentation masks, there is ample room for further research in localization and segmentation. Given that manifestations offer targeted feature descriptors for lump areas, diagnoses based on lump regions could achieve higher precision, albeit contingent on preceding detection or segmentation models.

5) The manifestation-driven reasoning, particularly integrating probabilistic graphical models, like Bayesian networks, with deep learning for intelligent inference models that more closely align with the reasoning and decision-making processes of imaging experts.

ViKL paves the way for future studies that integrate data-driven and knowledge-driven approaches in medical AI, offering extensive opportunities for advancing the field.

\section{Conclusion}
We have introduced ViKL, an innovative multimodal contrastive learning framework that harmonizes three critical elements of radiology practice: vision, language, and knowledge. Each modality contributes its unique strengths, particularly in terms of information density and flexibility, collectively enriching the diagnostic accuracy and enhancing the generalization potential of the framework in various downstream tasks. A noteworthy aspect of ViKL is its training approach, which necessitates only paired radiological modalities without the need for pathology labels. This feature is particularly advantageous, as acquiring large datasets with pathology labels is often a challenging and resource-intensive endeavor. 

Benefiting from its capability to construct a high-quality representation space, ViKL effectively calibrates the model's accuracy and confidence, significantly reducing the risk of over-confidence. Additionally, ViKL achieves a feature space characterized by excellent alignment and uniformity. This well-structured representation space is fundamental to ViKL's success in downstream tasks and its ability to generalize across different datasets. The robustness and adaptability of ViKL, driven by its advanced representation learning, demonstrate its potential as a versatile tool in medical image analysis, capable of providing reliable and accurate insights across a variety of scenarios.

Looking forward, we see tremendous potential for ViKL to spur further research in this domain. We encourage the research community to utilize the MKVL dataset to explore and expand the boundaries of multimodal learning. We envision this work contributing significantly to the advancement of medical AI, with the ultimate goal of improving healthcare outcomes. ViKL thus stands as a testament to the power of integrating diverse data modalities and a beacon for future innovations in medical imaging and analysis.
\ifCLASSOPTIONcaptionsoff
  \newpage
\fi

\footnotesize
\bibliographystyle{IEEEtranN}
\bibliography{refs}

\newpage
\appendices
\section{Hyperparameters}

\renewcommand{\arraystretch}{1.3}
\begin{table}[h]
    \centering
    \setlength{\tabcolsep}{7pt}
    \caption*{Hyperparameters. }
    \begin{threeparttable}
    \begin{tabular}{l c}
        \hline
        \hline
        \textbf{Hyperparameters} & \textbf{Value} \\
        \hline
        \multicolumn{2}{c}{Pretraining}\\
        \hline
        Peak learning rate & 1e-4 \\
        Minimal learning rate & 1e-7 \\
        Learning rate schedule & Cosine annealing \\
        Training steps &  9000 \\
        Warmup steps &  300 \\
        Batch size &  64 \\
        Temperature $\tau_1$ for img-mani and img-intra  &  0.03 \\
        Temperature $\tau_2$ for img-text and mani-text  &  0.3 \\
        Weight decay & 1e-4\\
        Input resolution & $256^{2}$\\
        ManiNeg, maximum $\mu$& 11\\
        ManiNeg, minimum $\mu$& 0\\
        ManiNeg, $\sigma$&3\\
        ManiNeg, annealing schedule&Linear\\
        ManiNeg, annealing, step at minimum $\mu$\tnote{$\dagger$}&50th\\

        \hline
        \multicolumn{2}{c}{Downstream, linear evaluation}\\
        \hline
        Peak learning rate &  1e-3 \\
        Minimal learning rate &  1e-6 \\
        Learning rate schedule &  Cosine annealing \\
        Maximum epochs &  1000 \\
        Early stopping patience & 100 \\
        Batch size &  48 \\
        Weight decay & 1e-6\\
        Input resolution & $256^{2}$\\
        \hline
        \multicolumn{2}{c}{Downstream, fine-tuning}\\
        \hline
        Peak learning rate  & 5e-5 \\
        Minimal learning rate &  5e-7 \\
        Learning rate schedule &  Cosine annealing \\
        Layer-wise learning rate decay\tnote{$\ddagger$}  &  0.1 \\
        Maximum epochs &  1000 \\
        Early stopping patience & 100 \\
        Batch size &  48 \\
        Weight decay & 5e-5\\
        Input resolution & $256^{2}$\\
        \hline
        \multicolumn{2}{c}{Downstream, linear probe}\\
        \hline
        $\mathcal{L}_2$ regularization strength $\lambda$  & 3.16 \\
        Maximum iteration &  1000 \\
        \hline
        \hline
    \end{tabular}
    
     \begin{tablenotes}
        \footnotesize
        \item[$\dagger$] During the training process, $\mu$ starts from a maximum value of 11 and stops at a minimum value of 0 at the 50th training step, remaining unchanged thereafter. The decrease in $\mu$ is linear.
        \item[$\ddagger$] The learning rate on the pretrained parameters is 0.1 times that of the learning rate on the appended randomly initialized linear layer.
      \end{tablenotes}
    \end{threeparttable}
\label{tab:hyper}
\end{table}
\newpage

\section{Detailed Network Structure}
\begin{figure}[h]
    \centering
    \includegraphics[width=0.49\textwidth]{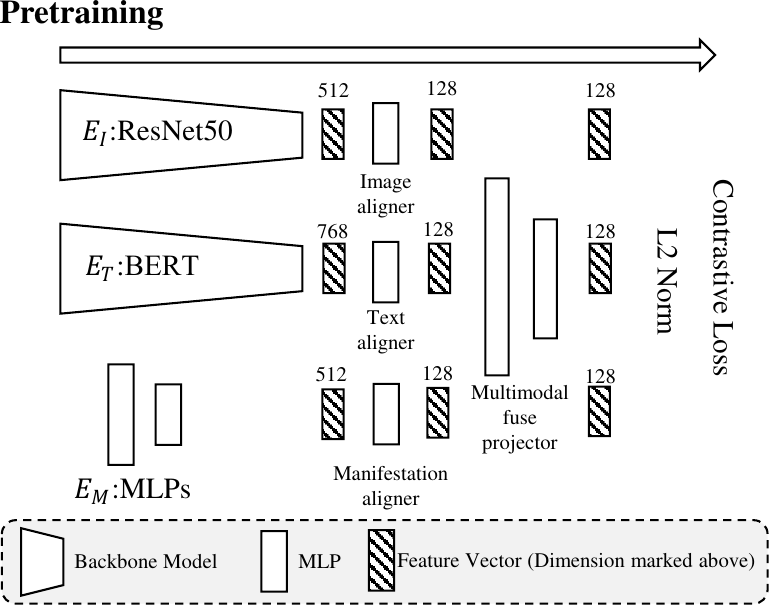} 
    \caption*{Detailed network structure of ViKL. For downstream tasks, the $E_I$ branch will be utilized, and a linear layer will be added as required by the specific evaluation protocol for benign-malignant classification.}
    \label{fig:appendixb}
\end{figure}
\newpage

\end{document}